\documentclass[aps,
prd,
preprint,
eqsecnum,
amsmath,
amssymb
]{revtex4}

\usepackage{hyperref}
\usepackage{mathrsfs}

\begin{document}

\title{Gravitational anomalies on the Newton-Cartan background} 

\author{Karan Fernandes} \email[email: ]{karan12t@bose.res.in}
\affiliation{S N Bose National Centre for Basic Sciences, Block JD, Sector III, Salt Lake, Kolkata 700106, India}

\author{Arpita Mitra} \email[email: ]{arpita12t@bose.res.in}
\affiliation{S N Bose National Centre for Basic Sciences, Block JD, Sector III, Salt Lake, Kolkata 700106, India}

\begin{abstract}
We derive the trace and diffeomorphism anomalies of the Schr\"odinger field minimally coupled to the Newton-Cartan background using Fujikawa's path integral approach. This approach in particular enables us to calculate the one-loop contributions due to all the fields of the Newton-Cartan structure. We determine the coefficients and demonstrate that gravitational anomalies for this theory always arise in odd dimensions. Due to the gauge field contribution of the background we find that in $2+1$ dimensions the trace anomaly contains terms which have a form similar to that of the $1+1$ and $3+1$ dimensional relativistic trace anomalies. This result reveals that the Newton-Cartan background which satisfies the Frobenius condition possesses a Type A trace anomaly in contrast with the result of Lishitz spacetimes. As an application we demonstrate that the coefficient of the term similar to the $1+1$ dimensional relativistic trace anomaly satisfies a c-theorem condition.
\end{abstract}


\maketitle

\section{Introduction}

Classical relativistic conformal theories coupled to curved backgrounds admit a stress-energy tensor which is symmetric, traceless and conserved. 
In contrast, quantum fields on curved backgrounds in general have a stress-energy tensor which violate these symmetries, resulting in gravitational anomalies  \cite{Bertlmann:1996xk,Fujikawa:2004cx,Duff:1993wm}. In considering relativistic systems with a symmetric stress-energy tensor, the trace anomaly arises when the quantum stress-energy tensor is not traceless, while its failure to be conserved results in the diffeomorphism anomaly. These anomalies have important consequences in black holes physics and cosmology \cite{Christensen:1977jc, Davies:1976ei,Hawking:1976ja,Candelas:1980zt,Jacobson:1991gr,Callan:1992rs,Robinson:2005pd,Solodukhin:2011gn,Davies:1977ze,Dowker:1975tf,Christensen:1979iy,Grishchuk:1981xf,delAguila:1986ea,Barvinsky:1992dz}, as well as in the computation of transport coefficients and response functions of condensed matter systems \cite{Stone:2012cx,Stone:2012ud,Gromov:2014gta,Gromov:2014vla,Gromov:2014dfa,Can:2014awa,Can:2014ota,Nishioka:2015uka,Iqbal:2015vka,Hughes:2015ora,Gromov:2015fda}. Gravitational anomalies are in addition background dependent, as evident from the difference of Lifshitz anomalies from those of relativistic backgrounds. Motivated by the extension of these results to nonrelativistic systems on curved backgrounds, we will be concerned with the trace and diffeomorphism anomalies of the Schr\"odinger field on the Newton-Cartan (NC) background. Note that while the trace anomaly of the NC background has been considered in \cite{Jensen:2014hqa,Jain:2015jla,Auzzi:2015fgg,Auzzi:2016lxb,Pal:2017ntk} following the discrete light-cone quantization (DLCQ) technique from higher dimensional relativistic backgrounds, our aim is to revisit the derivation starting from an action on the NC background. 
The interesting outcome of our derivation for the trace anomaly in 2+1 dimensions is that it takes the following general form
%
\begin{equation}
\left \langle 2 \widetilde{T}^{0}_{\phantom{\mu} 0} + \widetilde{T}^{i}_{\phantom{\mu} i} \right \rangle = \frac{1}{m (4\pi)^2}\left(\frac{1}{360} (R_{\mu \nu}h^{\mu \nu})^2  +  2 m^4 \psi^2 + \frac{m^2}{3} (\psi R_{\mu \nu}h^{\mu \nu} + R_{\mu \nu}v^{\mu} v^{\nu})\right)
\label{int.res}
\end{equation}
where $\psi = \tau^{\mu}A_{\mu} - \frac{1}{2}h^{\mu \nu}A_{\mu}A_{\nu}$ and $v^{\mu} = \tau^{\mu} - h^{\mu \nu}A_{\nu}$. The non-curvature squared terms of Eq.~(\ref{int.res}) were absent in the literature.

We will now briefly discuss the main results from prior considerations of the NC trace anomalies. Beginning with \cite{Jensen:2014hqa},  the trace anomaly was described as those terms in the most general Weyl variation which satisfy the Wess-Zumino consistency condition. In 2+1 dimensions, this was shown to be of the form of the 3+1 dimensional relativistic trace anomaly. It was further argued that the anomaly only arises in odd dimensions. In \cite{Jain:2015jla}, following the null background construction of \cite{Banerjee:2015hra,Banerjee:2015uta}, the anomaly was shown to be present in the same number of dimensions as relativistic theories. In \cite{Auzzi:2015fgg}, the trace anomaly was also demonstrated to arise in odd dimensions, following the embedding of the NC background in a relativistic background of one higher dimension \cite{Duval:1984cj}. The form of the anomaly in 2+1 dimensions was shown to be that of the 3+1 dimensional relativistic trace anomaly. The result of \cite{Auzzi:2015fgg} was rederived in \cite{Auzzi:2016lxb} using a heat kernel approach. 
The results of \cite{Auzzi:2015fgg,Auzzi:2016lxb} as well as our own are in disagreement with that of \cite{Pal:2017ntk}.

Non-relativistic anomalies can receive contributions due to $c^{-1}$ and $m$ corrections of relativistic field theories, resulting in terms unlike those in the relativistic theory. This can be particularly appreciated through the derivation of the non-relativistic scale anomaly in \cite{Bergman:1991hf}. Such subtleties are best addressed within effective field theory approaches. As mentioned in the literature, the metric structure of the NC background introduces several obstacles.  This background possesses two mutually orthogonal, degenerate metrics and an additional gauge field $A_{\mu}$ \cite{D}. Not only are there more than one metric, but their variations must satisfy certain relations among themselves to maintain the NC structure. While this leads to several interesting consequences for fields coupled to them, it significantly complicates the computation of gravitational anomalies. The heat kernel approach in \cite{Auzzi:2015fgg} describes some of these complications in the process of deriving its results about flat space, where in addition the gauge field $A_{\mu}$ was set to vanish. As we will describe, this gauge field is central to the result we derive. 

While several techniques may be employed in the calculation of anomalies, we have found Fujikawa's method \cite{Fujikawa:2004cx, Fujikawa:1979ay, Fujikawa:1980rc} particularly appropriate given the NC background. Fujikawa's approach recognizes the anomaly as the failure of the measure of the path integral to remain invariant under the given symmetry transformation. One of the ways to evaluate the functional trace of the Jacobian for gravitational anomalies is through a regulator and basis, for which we will use the plane wave approach of \cite{Ceresole:1988hn,Hatsuda:1989qy}. This approach leads to the correct result for the relativistic trace, chiral and diffeomorphism anomalies. The regulator was introduced in \cite{Lindstrom:1988fr}, which was further shown to be equivalent to Pauli-Villars regularization in \cite{Diaz:1989nx}. Evaluating the trace of the regulated Jacobian leads to candidate anomaly terms, not all of which are the true anomaly. The general expression contains terms for which a counterterm can be included in the effective action. Only those terms which cannot be written as a counterterm constitute the anomaly. The significant drawbacks of this approach are the Baker-Campbell-Haussdorff (BCH) expansion to high orders and the evaluation of large $k$ integrals, making them unfeasible for higher dimensional anomalies. Nevertheless, since the result follows only from a plane wave expansion and the variations of the Schr\"odinger fields, it turns out to be very useful for the NC background. 

We also plan on exploring the implications of the anomaly terms of Eq.~(\ref{int.res}) in the renormalization group (RG) flow of the corresponding fields. It is well known that relativistic trace anomalies impose non-trivial constraints on the infrared dynamics emerging from an ultraviolet unitary theory. 
These constraints follow from imposing the Wess-Zumino (WZ) consistency
conditions on the local Callan-Symanzik (CS) equation, which in 2 dimensions provides a proof of the Zamolodchikov c-theorem \cite{Jack:1990eb}. By relying entirely on the Abelian nature of Weyl transformations and the general form of the anomaly density, this procedure can provide a non-perturbative proof  without the requirement of any particular renormalization scheme. The formulation of the consistency conditions in $z=2$ theories has been considered in \cite{Pal:2016rpz}. The consistency conditions which result from the NC anomaly terms was studied in \cite{Auzzi:2016lrq}. We will demonstrate how the Weyl consistency condition implies that the term $R_{\mu \nu}\tau^{\mu}\tau^{\nu}$ contained in Eq.~(\ref{int.res}) satisfies a c-theorem condition. To demonstrate that the anomaly coefficient satisfies a definite monotonicity property, we would need to consider the correlation functions of the Schr\"odinger fields. As the derivation of these correlators lies beyond the scope of the present work, we will address this in the future. 

The organization of our paper is as follows. In Sec.~\ref{NC}, we review basic properties of the NC background which will be relevant to our derivation. In Sec.~\ref{Schr}, we consider the Schr\"odinger action on the NC background and its symmetries.  In Sec.~\ref{anom}, we derive the diffeomorphism and trace anomalies using Fujikawa's approach in a plane-wave basis. In Sec.~\ref{cthm}, we demonstrate that the coefficient of the $R_{\mu \nu}\tau^{\mu}\tau^{\nu}$ anomaly term satisfies a c-theorem condition. Finally, in Sec.~\ref{sec.con} we conclude with a discussion of our result and their implications for systems on curved Newtonian backgrounds. Appendix~\ref{app.ac} provides details of the adapted coordinate system for the NC background, which is used to calculate the anomalies. Appendix~\ref{app2.app2} reviews Fujikawa's approach, the regulator used in relativistic field theories and the regulator used in this work for non-relativistic field theories. Appendix~\ref{app.C} contains intermediate details needed for the calculation provided in Sec.~\ref{anom}. \\

\section{The Newton-Cartan background} \label{NC}


The NC background was initially constructed by Cartan in~\cite{Cartan:1923zea}, as a covariant spacetime formulation of Newtonian gravity. Further investigations detailed the geometric properties of the background~\cite{D,Kunzle72}, in particular its relation to the Bargmann algebra~\cite{Bargmann:1954gh} and the minimal coupling of fields to it~\cite{Duval:1976ht, Duval:1983pb, Kunzle:1984dt}. This background has been subsequently derived in a number of ways including the reduction from a higher dimensional relativistic background~\cite{Duval:1984cj,Kunzle:1985bj}, the gauging of the Bargmann algebra~\cite{Andringa:2010it,Bergshoeff:2014uea}, coset construction~\cite{Brauner:2014jaa, Jensen:2014aia} and the localization of spacetime symmetries of the Schr\"odinger field~\cite{Banerjee:2014nja,Banerjee:2015rca,Mitra:2015twa}.  We will here review certain properties of the metric and connection of the torsion-free NC background relevant for later sections. 

The NC background contains a degenerate inverse spatial metric and a degenerate temporal 1-form satisfying the following relations,
\begin{align}
\nabla_{\mu}h^{\alpha \beta} = 0 \qquad & \qquad \nabla_{\mu}\tau_{\nu} = 0 \label{nc.met} \\
h^{\mu \nu} \tau_{\mu} &= 0
\label{nc.orth}
\end{align}
Given that $h^{\mu \nu}$ and $\tau_{\mu}$ are degenerate, their inverses do not exist. We can formally define a generalized inverse for $\tau^{\mu}$ such that
\begin{equation}
\tau^{\mu}\tau_{\mu} = 1
\label{nc.norm}
\end{equation}
We can further define a spatial metric $h_{\mu \nu}$ that satisfies the following relations
\begin{align}
h_{\mu \nu}\tau^{\mu} &= 0 \, , \notag \\
h^{\mu \lambda}h_{\lambda \nu} + \tau^{\mu}\tau_{\nu} &= \delta^{\mu}_{\nu} \, .
\label{nc.proj}
\end{align}
Unlike $h^{\mu \nu}$, the covariant derivative of $h_{\mu \nu}$ does not vanish. The variation of $h_{\mu\nu}$ follows from Eq.~(\ref{nc.proj}),
\begin{equation}
\delta h_{\mu\nu}=-2h_{\rho(\mu}\tau_{\nu)}\delta \tau^{\rho} \, .
\end{equation}
Thus variations and derivatives of $h_{\mu\nu}$ are not independent of $\tau^{\mu}$ and we must choose either $h_{\mu \nu}$ or $\tau^{\mu}$ as the independent field. Conventionally $\tau^{\mu}$ is taken to be the independent field, which will also be followed in this paper. 

A direct consequence of the metricity conditions is that the connection is not uniquely determined by these metrics alone. The most general linear, symmetric connection which satisfies Eq.~(\ref{nc.met}) has the form 
\begin{equation}
{\Gamma^\rho}_{\nu\mu} = \tau^{\rho}\partial_{(\mu}\tau_{\nu)} +
\frac{1}{2}h^{\rho\sigma} \Bigl(\partial_{\mu}h_{\sigma\nu}+\partial_{\nu}h_{\sigma\mu} - \partial_{\sigma}h_{\mu\nu}\Bigr)+ h^{\rho\lambda}\tau_{(\mu}K_{\nu) \lambda} \, , 
\label{nc.con}
\end{equation}
where $K_{\lambda \mu}$ is just an arbitrary two form at this stage.
One can now construct the Riemann tensor for a symmetric connection in the usual way  
\begin{equation}
[\nabla_{\mu}, \nabla_{\nu}]V^{\lambda}=R^{\lambda}_{\phantom{\lambda}\sigma\mu\nu}V^{\sigma}\label{R}
\end{equation}
where $R^{\lambda}{}_{\sigma\mu\nu}$ satisfy the following relations,
\begin{equation}
\tau_{\rho}R^{\rho}_{\phantom{\rho}\sigma\mu\nu}=0,~~R^{\lambda}_{\phantom{\lambda}\sigma(\mu\nu)}=0,~~R^{\lambda}_{\phantom{\lambda}[\sigma\mu\nu]}=0,
~~R^{(\lambda\sigma)}{}{}_{\mu\nu}=0
\label{nc.RSsymm}
\end{equation}
The NC connection can be demonstrated as the Newtonian limit of the connection of a Riemannian manifold provided Trautman's condition holds \cite{Duval:1983pb}
\begin{equation}
R^{\lambda}{}_{\sigma}{}^{\mu}{}_{\nu}=R^{\mu}{}_{\nu}{}^{\lambda}{}_{\sigma} 
\label{nc.trautman}
\end{equation}
Indices were raised in Eq.~(\ref{nc.trautman}) using the metric $h^{\mu \nu}$. From Eq.~(\ref{nc.con}) we note that Eq.~(\ref{nc.trautman}) is equivalent to requiring $dK=0$. This implies that 
\begin{equation}
K_{\lambda \mu} = 2 \partial_{[\lambda} A_{\mu]} \, , \label{nc.gauge}
\end{equation}
where $A_{\mu}$ is an arbitary 1-form.
Non-relativistic spacetimes also do not have a preferred vector field $\tau^{\mu}$ and this leads to an additional invariance under Milne boosts \cite{Jensen:2014aia,Geracie:2015dea} which are described by
 \begin{align}
\tau^{\mu} &\rightarrow \tau^{\mu}+h^{\mu\nu}k_{\nu}\, , \notag\\
h_{\mu \nu} & \rightarrow h_{\mu \nu} - 2\tau_{(\mu} k_{\nu)} + \tau_{\mu} \tau_{\nu} h^{\alpha \beta}k_{\alpha} k_{\beta}\, , \notag\\
A_{\mu} & \rightarrow A_{\mu} + k_{\mu} -\frac{1}{2}\tau_{\mu}h^{\alpha \beta}k_{\alpha} k_{\beta}\, .
 \end{align} 
where $k_{\mu}$ is an arbitrary spatial vector, i.e. $k^{\mu}\tau_{\mu}=0$. The NC background and its torsion free connection Eq.~(\ref{nc.con}) are invariant under this transformation.
%

A covariant measure for the NC background follows by defining the nowhere vanishing effective metric $\gamma_{\mu \nu} = h_{\mu \nu} + \tau_{\mu} \tau_{\nu}$~\cite{Jensen:2014aia}. While this metric is neither Milne invariant nor does it satisfy the metricity condition, its determinant satisfies both. Given a $2+1$ dimensional NC spacetime which satisfies the Frobenius condition, we can describe the determinant as
\begin{align}
\vert \gamma \vert & = \frac{1}{3!} \epsilon^{\mu \nu\sigma} \epsilon^{\alpha \beta \gamma}\gamma_{\mu \alpha}\gamma_{\nu \beta} \gamma_{\sigma \gamma} \notag\\
& = \vert h \vert \, ,
\label{detmet}
\end{align}
The second equality of Eq.~(\ref{detmet}) follows from Eq.~(\ref{nc.proj}) and the unit lapse function of the NC spacetime (Eq.~(\ref{nc.norm})). Thus the measure of the NC background is simply given by $\sqrt{h}$ when it satisfies the Frobenius condition \footnote{This is often noted in the literature as $\sqrt{\vert h_{\mu \nu} + \tau_{\mu}\tau_{\nu} \vert}$. This is indeed the correct measure to use in the case of NC backgrounds with torsion and cannot be simplified as was done in Eq.~(\ref{detmet}). When the Frobenius condition is satisfied however, we can use the same derivation as that considered within the ADM formulation of General Relativity. Also, just as in the ADM formulation, while $h_{\mu \nu}$ is a degenerate spacetime metric, $h_{ij}$ is a nondegenerate spatial metric of the hypersurface. It is the determinant of this metric which enters in the ADM measure $\lambda \sqrt{h}$, where $\lambda$ is the lapse function.} .


The relations considered in this section are valid for the symmetric connection of the NC background. The presence of torsion leads to a NC background for which $d\tau \neq 0$. In addition, such backgrounds require the construction of a modified connection in order to ensure invariance under both Milne and $U(1)$ transformations. For further details on NC backgrounds with torsion we refer the reader to \cite{Bergshoeff:2014uea,Jensen:2014aia}. All the calculations in this paper will be considered on the NC background without torsion.\\

\section{The Schr\"odinger field on the NC background} \label{Schr}

The Schr\"odinger field on the NC background was originally considered in \cite{Duval:1983pb}, with the intent of providing the (Galilean) covariant Schr\"odinger equation on curved Newtonian backgrounds. More recently, this action has received attention due to its many newfound applications in condensed matter physics \cite{Banerjee:2015rca,Son:2005rv,Wu:2014dha} and holography \cite{Hartong:2014oma,Christensen:2013lma}. In $2+1$ dimensions, this action can be written as,
\begin{align}
S &= \int dt d^2x \sqrt{h} \mathcal{L} \notag\\
&= \int dt d^2x \sqrt{h} \left[ i m\left(\Phi^* \tau^{\mu} \mathcal{D}_{\mu}\Phi - \Phi \tau^{\mu} \bar{\mathcal{D}}_{\mu}\Phi^* \right) - h^{\mu \nu}\mathcal{D}_{\mu}\Phi \bar{\mathcal{D}}_{\nu}\Phi^* \right] \, ,
\label{sf.act}
\end{align}
where $\mathcal{D}_{\mu} = \nabla_{\mu} - i m A_{\mu}$, $\bar{\mathcal{D}}_{\mu} = \nabla_{\mu} + i m A_{\mu}$ and $\nabla_{\mu}$ represents the usual covariant derivative of the spacetime. The gauge field $A_{\mu}$ is a mass generating field which provides particle number conservation on the NC background. It is also the same field which appears in the NC connection contained in $\nabla_{\mu}$ and is therefore on the same footing as all other gravitational fields. In addition, the action Eq.~(\ref{sf.act}) is known to be invariant under Milne boosts \cite{Jensen:2014aia}. In this regard, it will be useful to define the Milne invariant quantities
\begin{align}
v^{\mu} &= \tau^{\mu} - h^{\mu \nu}A_{\nu}  = \tau^{\mu} - A^{\mu} \notag\\
\psi & = \tau^{\mu}A_{\mu} - \frac{1}{2}h^{\mu \nu}A_{\mu}A_{\nu}
\label{sf.mil}
\end{align}  
For convenience we will also define $\partial^{\mu} = h^{\mu \nu}\partial_{\nu}$. Note that in Eq.~(\ref{sf.act}), $m$ is merely a passive parameter with no mass dimension \cite{Bergman:1991hf}. 

Since we are interested in understanding the symmetries of Eq.~(\ref{sf.act}), let us first consider its total variation 
\begin{equation}
\delta S = \int dt d^2x \sqrt{h} \left[ - P_{\mu \nu} \delta h^{\mu \nu} + R_{\mu} \delta \tau^{\mu} - J^{\mu} \delta A_{\mu} + \delta \Phi^* \mathcal{D} \Phi + \delta \Phi \bar{\mathcal{D}} \Phi^* \right] \, ,
\label{sf.gvar}
\end{equation}
where we have defined
\begin{align}
P_{\mu \nu} &= \frac{1}{2}h_{\mu \nu} \mathcal{L} + \mathcal{D}_{\mu}\Phi \bar{\mathcal{D}}_{\nu}\Phi^* \notag\\
R_{\mu} &= im \left(\Phi^* \mathcal{D}_{\mu}\Phi - \Phi \bar{\mathcal{D}}_{\mu}\Phi^* \right) \notag\\
J^{\mu} &= - 2 m^2 \Phi \Phi^* v^{\mu} + im\left( \Phi^* \partial^{\mu}\Phi - \Phi \partial^{\mu}\Phi^*\right) \notag\\
\mathcal{D} \Phi &= \left(2 im v^{\mu}\nabla_{\mu} + im \nabla_{\mu}v^{\mu} + 2 m^2 \psi + h^{\mu \nu} \nabla_{\mu} \nabla_{\nu}\right) \Phi \notag\\
\bar{\mathcal{D}} \Phi^* &= \left(- 2 im v^{\mu}\nabla_{\mu} - im \nabla_{\mu}v^{\mu} + 2 m^2 \psi + h^{\mu \nu} \nabla_{\mu}\nabla_{\nu}\right) \Phi^*
\label{sf.coeff}
\end{align}
We note again that as variations of $\delta h_{\mu \nu}$ are not independent of $\delta \tau^{\mu}$, they do not appear separately in Eq.~(\ref{sf.gvar}). Let us now consider the variations to be diffeomorphisms with respect to some arbitrary vector field $\xi^{\mu}$, i.e. $\delta_{\xi} = \pounds_{\xi}$, with $\pounds$ denoting the Lie derivative. It is straightforward to demonstrate $\delta_{\xi} S = 0$ and hence Eq.~(\ref{sf.act}) is invariant under diffeomorphisms. 

We further consider the on-shell symmetries of the action
\begin{align}
0 = \delta_{\xi} S &= \int dt d^2x \sqrt{h}  [-P_{\mu\nu}{\pounds}_{\xi} h^{\mu\nu}  + R_{\mu}{\pounds}_{\xi} \tau^{\mu} - J^{\mu}{\pounds}_{\xi} A_{\mu}]   \notag\\
&= \int dt d^2x 2 \sqrt{h} \xi^{\nu}[ -\nabla_{\mu}T^{\mu}{}_{\nu} - J^{\mu}\nabla_{[\nu}A_{\mu]} + \frac{1}{2}R_{\mu}\nabla_{\nu}\tau^{\mu}]
\label{sf.diff}
\end{align}
Here $T^{\mu}{}_{\nu}$ is the stress tensor of the Schr\"odinger field on the NC background, which is defined as
\begin{equation}
T^{\mu}{}_{\nu} =  P_{(\nu \sigma)} h^{\sigma \mu} - \frac{1}{2}R_{\nu} \tau^{\mu} \, .
\label{sf.st}
\end{equation}
Thus Eq.~(\ref{sf.act}) remains invariant under on-shell diffeomorphisms provided the stress tensor satisfies
\begin{equation}
\nabla_{\mu}T^{\mu}{}_{\nu} + J^{\mu}\nabla_{[\nu}A_{\mu]} - \frac{1}{2} R_{\mu}\nabla_{\nu}\tau^{\mu} = 0 \, .
\label{sf.con1}
\end{equation}

Let us now consider Weyl transformations, $\delta_{\Lambda} = w \Lambda$, where $w$ is the weight of the field and $\Lambda$ is the parameter of the transformation. It can be noted that the action Eq.~(\ref{sf.act}) is not Weyl invariant ($\delta_{\Lambda} S \neq 0$) and thus cannot be used to investigate the Weyl anomaly.
Usually one could now include a term proportional to $R \Phi {\Phi}^*$  and determine the proportionality constant which ensures invariance. However, in $2+1$ dimensions we can construct a Weyl-invariant action from Eq.~(\ref{sf.act}) by replacing the scalar fields with scalar densities. This trick is known to work for relativistic scalar fields in $1+1$ dimensions, where the densitized fields are known as Fujikawa variables.

By substituting $\Phi = \widetilde{\Phi} h^{-\frac{1}{4}}$ and $\Phi^* = \widetilde{\Phi}^* h^{-\frac{1}{4}}$ in Eq.~(\ref{sf.act}), we have
\begin{align}
\widetilde{S}  &= \int dt d^2x \sqrt{h} \widetilde{\mathcal{L}} \notag \\
&= \int dt d^2x \sqrt{h} \left[ i m h^{-\frac{1}{4}}\left(\widetilde{\Phi}^* \tau^{\mu} \mathcal{D}_{\mu}(\widetilde{\Phi} h^{-\frac{1}{4}}) - \widetilde{\Phi} \tau^{\mu} \bar{\mathcal{D}}_{\mu}(\widetilde{\Phi}^* h^{-\frac{1}{4}}) \right) \right. \notag\\
& \left. \qquad \qquad \qquad \qquad \qquad \qquad \qquad \quad - h^{\mu \nu}\mathcal{D}_{\mu}(\widetilde{\Phi} h^{-\frac{1}{4}}) \bar{\mathcal{D}}_{\nu}(\widetilde{\Phi}^* h^{-\frac{1}{4}}) \right]  
\label{sf.act2}
\end{align}
The fundamental fields of Eq.~(\ref{sf.act2}) are now $\{\widetilde{\Phi}, \widetilde{\Phi}^*, A_{\mu}, h^{\mu \nu}, \tau^{\mu}\, , \tau_{\mu} \}$. The total variation of the action Eq.~(\ref{sf.act2}) in this case can be expressed as
\begin{align}
\delta \widetilde{S} &= \int dt d^2x \left[ - \widetilde{P}_{\mu \nu} \delta h^{\mu \nu} +  \widetilde{R}_{\mu} \delta \tau^{\mu} -  \widetilde{J}^{\mu} \delta A_{\mu} + \delta \widetilde{\Phi}^*\mathscr{R} \widetilde{\Phi} + \delta \widetilde{\Phi} \left(\mathscr{R} \widetilde{\Phi}\right)^* \right]
\label{sf.gvar2}
\end{align}
with
\begin{align}
\widetilde{P}_{\mu \nu} &= \frac{\sqrt{h}}{2}h_{\mu \nu} \widetilde{\mathcal{L}} + \sqrt{h}\mathcal{D}_{\mu}(\widetilde{\Phi} h^{-\frac{1}{4}}) \bar{\mathcal{D}}_{\nu}(\widetilde{\Phi}^* h^{-\frac{1}{4}}) - \frac{1}{4} h_{\mu \nu} \left(\widetilde{\Phi}^* \mathscr{R}\widetilde{\Phi} + \widetilde{\Phi}(\mathscr{R}\widetilde{\Phi})^*\right) \notag\\
\widetilde{R}_{\mu} &= im h^{\frac{1}{4}}\left( \widetilde{\Phi}^* \mathcal{D}_{\mu} (\widetilde{\Phi}h^{-\frac{1}{4}}) - \widetilde{\Phi} \bar{\mathcal{D}}_{\mu} (\widetilde{\Phi}^* h^{-\frac{1}{4}}) \right) \notag\\
\widetilde{J}^{\mu} &= 2 m^2 \widetilde{\Phi}\widetilde{\Phi}^*v^{\mu} + imh^{\frac{1}{4}}\left( \widetilde{\Phi}^* \partial^{\mu}(\widetilde{\Phi}h^{-\frac{1}{4}}) - \widetilde{\Phi} \partial^{\mu}(\widetilde{\Phi}^*h^{-\frac{1}{4}})\right) \notag\\
\mathscr{R}\widetilde{\Phi} &=  \left(h^{\frac{1}{4}}\mathcal{D} h^{-\frac{1}{4}}\right)  \widetilde{\Phi} = \left[h^{\frac{1}{4}} \left(2 im v^{\mu}\nabla_{\mu} + im \nabla_{\mu}v^{\mu} + 2 m^2 \psi + h^{\mu \nu} \nabla_{\mu}\nabla_{\nu}\right)  h^{-\frac{1}{4}}\right] \widetilde{\Phi} \notag\\
\left(\mathscr{R}\widetilde{\Phi}\right)^* &=\left(h^{\frac{1}{4}}\bar{\mathcal{D}}h^{-\frac{1}{4}}\right) \widetilde{\Phi}^* = \left[h^{\frac{1}{4}} \left(- 2 im v^{\mu}\nabla_{\mu} - im \nabla_{\mu}v^{\mu} + 2 m^2 \psi + h^{\mu \nu} \nabla_{\mu}\nabla_{\nu}\right)  h^{-\frac{1}{4}}\right] \widetilde{\Phi}^* 
\label{sf.coeff2}
\end{align}
 
We now find that Eq.~(\ref{sf.gvar2}) vanishes under
\begin{align}
\delta_{\Lambda} \widetilde{\Phi} = \Lambda \widetilde{\Phi} \, , & \, \delta_{\Lambda} \widetilde{\Phi}^* = \Lambda \widetilde{\Phi}^* \, , \label{sf.fiwtran}\\
\delta_{\Lambda} h^{\mu \nu} = -2 \Lambda h^{\mu \nu} \, , \, \delta_{\Lambda} \tau^{\mu} &= -2 \Lambda \tau^{\mu} \, , \delta_{\Lambda} A_{\mu} = 0 \, . 
\label{sf.wtran}
\end{align}
Thus the action Eq.~(\ref{sf.act2}) is invariant under Weyl transformations. Considering the on-shell invariance of Eq.~(\ref{sf.act2}) under Weyl transformations, we find
 \begin{align}
0 = \delta_{\Lambda} \widetilde{S} &= \int dt d^2x \sqrt{h}  [-\widetilde{P}_{\mu\nu}\delta_{\Lambda} h^{\mu\nu}  + \widetilde{R}_{\mu}\delta_{\Lambda} \tau^{\mu}]  \notag\\
&= \int dt d^2x \sqrt{h} 2 \Lambda [ 2\widetilde{T}^{0}{}_{0} + \widetilde{T}^{i}{}_{i}]\, ,
\label{sf.scale}
\end{align}
where 
\begin{align}
2\widetilde{T}^{0}{}_{0} + \widetilde{T}^{i}{}_{i} &= - \frac{1}{2} (\widetilde{\Phi}^* \left(h^{-\frac{1}{4}} \mathcal{D} h^{-\frac{1}{4}}\right)  \widetilde{\Phi} + \widetilde{\Phi}  \left(h^{-\frac{1}{4}}\bar{\mathcal{D}}h^{-\frac{1}{4}}\right)  \widetilde{\Phi}^*) \, ; \notag\\
\widetilde{T}^{0}{}_{0} &:= \widetilde{T}^{\mu}{}_{\nu} \tau^{\nu}\tau_{\mu} \, , \quad \qquad  \widetilde{T}^{i}{}_{i} := \widetilde{T}^{\mu}{}_{\nu}h_{\mu \alpha} h^{\alpha \nu} \, .
\end{align}
It is evident from Eq.~(\ref{sf.scale}) that the on-shell Weyl invariance of Eq.~(\ref{sf.act2}) can be restored provided
\begin{equation}
2\widetilde{T}^{0}{}_{0} + \widetilde{T}^{i}{}_{i} = 0 \, .
\label{sf.con2}
\end{equation}
We have thus demonstrated that the $2+1$ dimensional Schr\"odinger field on the Newton-Cartan background can be used to investigate its invariance under both diffeomorphisms and Weyl transformations (the latter by densitizing the Schr\"odinger fields). This will be particularly useful in investigating both trace and diffeomorphism anomalies in the following section.\\

\section{Derivation of the gravitational anomalies}\label{anom}
The invariance of the path integral under the symmetries provided in the previous section leads to anomalous on-shell stress tensor relations. Specifically, the path integral average of the on-shell stress tensor relations of Eq.~(\ref{sf.con1}) and Eq.~(\ref{sf.con2}) are now equal to the functional trace of the Jacobian of the Schr\"odinger fields under the given symmetry transformation. For the derivation of gravitational anomalies, this trace is evaluated using an appropriate regulator ($\mathcal{R}$) and Jacobian ($J$). Following Eq.~(\ref{app2.regjac}), we can write the actions of the previous section as
\begin{equation} 
S = \int dt d^2x \frac{1}{2} \Psi^* \mathbf{T} \mathcal{R} \Psi \, ,
\label{an.pact}
\end{equation}
where $\Psi$ and $\Psi^*$ are the quantum fields (which may now be viewed as flat space fields), as all gravitational field dependence is now absorbed into the definitions of $\mathbf{T}$ and $\mathcal{R}$. In this case, given $\delta \Psi = K \Psi$ the Jacobian may be written as
\begin{equation}
J = K + \frac{1}{2}\mathbf{T}^{-1} \delta \mathbf{T}
\label{an.jaco}
\end{equation} 
A detailed review behind this choice is provided in Appendix \ref{app2.app2}. The gravitational anomaly now results from the following regulated trace
\begin{align}
A(x) &= \int dt d^2x ~\text{An}(x)\notag\\
\text{An}(x)& = \lim_{M \to \infty} \text{Tr}~ J e^{\frac{\mathcal{R}}{M^2}} \, .
\label{an.regtr}
\end{align} 
$\text{An}(x)$ refers to the anomaly (density) expressions we will derive in this work. To evaluate the trace in Eq.~(\ref{an.regtr}), we expand $\Psi$ and $\Psi^*$ as flat space plane wave modes so that the result follows from Gaussian integration. In the nonrelativistic case the regulated trace to be used is given by 
\begin{equation}
\lim_{M \to \infty} \text{Tr} J = \lim_{M \to \infty} \int \limits_{0}^{\infty}  \frac{d\omega}{2 \pi} \int \limits_{-\infty}^{\infty}\frac{d^2k}{(2 \pi)^2} e^{-i \omega t} e^{ikx}\left[J(x) e^{\frac{\mathcal{R}}{M^2}} \right] e^{i \omega t} e^{- ikx} \, .
\end{equation}
The reason behind the above integral representation is provided in \ref{app2.app3}. We will now derive the trace and diffeomorphism anomalies by evaluating this integral.\\

\subsection{The trace anomaly}
To derive the trace anomaly we consider the action Eq.~(\ref{sf.act2}), which can be expressed as
\begin{align}
S &= \int dt d^2x \widetilde{\Phi}^* \mathscr{R} \widetilde{\Phi} \, ,
\label{an.act1}
\end{align}
where $\widetilde{\Phi}$ and $\widetilde{\Phi}^*$ are the fundamental fields and $\mathscr{R}$ is the Hermitian operator defined in Eq.~(\ref{sf.coeff2}). The path integral is given by 
\begin{equation}
Z = \int \mathcal{D}\widetilde{\Phi} \mathcal{D}\widetilde{\Phi}^* e^{i S\left[\widetilde{\Phi}, \widetilde{\Phi}^*, \tau^{\mu}, h^{\mu\nu}, A_{\mu}\right]} \, .
\label{an.pi1}
\end{equation}
Using Eq.~(\ref{sf.fiwtran}), we find that the invariance of Eq.~(\ref{an.pi1}) under Weyl transformations of the fields $\widetilde{\Phi}$ and $\widetilde{\Phi}^*$  results in the following anomalous Ward identity
\begin{align}
\left \langle  \Lambda \sqrt{h} \left(2\widetilde{T}^{0}{}_{0} + \widetilde{T}^{i}{}_{i} \right) \right \rangle_{\widetilde{\Phi}\widetilde{\Phi}^*} &= \left \langle \text{Tr} J \right \rangle_{\widetilde{\Phi}\widetilde{\Phi}^*} \, ,
\label{an.an}
\end{align}
where $\langle\cdots \rangle_{\widetilde{\Phi} \widetilde{\Phi}^*}$ denotes the path integral average with respect to the variables $\widetilde{\Phi}$ and $\widetilde{\Phi}^*$. To proceed, we regulate the trace occuring in Eq.~(\ref{an.an}) 
\begin{align}
\left \langle \text{Tr} J \right \rangle_{\widetilde{\Phi}\widetilde{\Phi}^*} \to \lim_{M \to \infty}\text{Tr} J e^{\frac{\mathcal{R}}{M^2}}   \, .
\label{an.an1}
\end{align}
The Jacobian and the regulator to be used can be determined by comparing Eq.~(\ref{an.act1}) with Eq.~(\ref{an.pact}) and Eq.~(\ref{an.jaco}). The Jacobian is simply $J= \Lambda(x)$ (since $\mathbf{T} = 2$) while the regulator is
\begin{equation}
\mathcal{R} = \mathscr{R} = h^{\frac{1}{4}} \mathcal{D} h^{-\frac{1}{4}}  \, , 
\label{an.reg1}
\end{equation}
The regulated trace which needs to be evaluated is now given by
\begin{equation}
\lim_{M \to \infty}\text{Tr}\Lambda(x) e^{\frac{\mathcal{R}}{M^2}}  = \lim_{M \to \infty} \int \limits_{0}^{\infty}  \frac{d\omega}{2 \pi} \int \limits_{-\infty}^{\infty} \frac{d^2k}{(2 \pi)^2} e^{-i \omega t} e^{ikx}\left[\Lambda(x) e^{\frac{\mathcal{R}}{M^2}} \right] e^{i \omega t} e^{- ikx} \, .
\label{an.wtr}
\end{equation}
Due to the use of flat space nonrelativistic plane waves we expand $\mathcal{R}$ in the adapted coordinates described in \ref{app.ac}. For the calculation to follow it will be useful to decompose the Milne invariant quantities in Eq.~(\ref{sf.mil}) as $v^{\mu} = \left \{v^0, v^i\right\}$ and $\psi = \phi + \bar{\phi}$, where
\begin{align}
v^0 &=\tau^0, ~v^i =\tau^i - h^{ij}A_{j} \, ,  \notag\\
\phi &= \tau^0 A_{0}, ~\bar{\phi} = \tau^{i} A_{i} - \frac{1}{2} h^{i j}A_{i}A_{j} \, .
\label{an.milne}
\end{align}  
With these definitions Eq.~(\ref{an.reg1}) can be written as
\begin{equation}
\mathcal{R} = h^{\frac{1}{4}} \left[2 i m v^0 \partial_t + 2 i m v^i \partial_{i} + h^{i j} \left(\partial_{i} \partial_{j} - \Gamma_{i j}^{k}\partial_{k} \right) - i m \mathcal{C}\right]h^{-\frac{1}{4}} \, ,
\label{an.regm}
\end{equation} 
where $\partial_t = \frac{\partial}{\partial t}$ and $\mathcal{C}$ are given by
\begin{equation}
\mathcal{C}= -\nabla_{i}v^i + 2im\left(\bar{\phi} + \phi\right) \, .
\label{an.regc}
\end{equation}
We can now move the plane wave $e^{i \omega t} e^{- ikx}$ from the right of the regulator in Eq.~(\ref{an.wtr}) to the left. By further rescaling $k \to M k$ and $\omega \to M^2 \omega$ we have
\begin{equation}
\lim_{M \to \infty}\text{Tr}\Lambda(x) e^{\frac{\mathcal{R}}{M^2}} = \lim_{M \to \infty} M^4 \int  \frac{d\omega}{2 \pi} \int  \frac{d^2 k}{(2\pi)^2} \Lambda(x) e^{\frac{\mathcal{R}(M k, M^2 \omega)}{M^2}} \, ,
\label{an.ltor}
\end{equation}
where the operator in the exponent now takes the form
\begin{align}
\frac{\mathcal{R}(M k, M^2 \omega)}{M^2} &= -k^2 - 2 m v^0 \omega + \frac{1}{M} \left(i k_i \Gamma^i - 2 i k_i \partial^i + 2 m k_i v^i  - 2ih^{\frac{1}{4}}k_i\partial^i(h^{-\frac{1}{4}})\right) \notag\\
&\qquad \qquad \qquad +\frac{1}{M^2}\left(\Delta - i m \mathcal{C}  + h^{\frac{1}{4}}\Delta h^{-\frac{1}{4}} + 2 h^{\frac{1}{4}} \partial^l (h^{-\frac{1}{4}}) \partial_l\right) \, . 
\label{an.regkw}
\end{align}
In Eq.~(\ref{an.regkw}) we have used the following definitions,
\begin{align}
\Gamma^i &= h^{mn} \Gamma^i_{mn} \, ,  \quad k^2 = k_i k_j h^{ij} \, ,\notag\\
\Delta &= \partial^i\partial_j -\Gamma^i\partial_i + 2imv^0\partial_t + 2imv^i\partial_i \, .
\end{align}
At this stage we can factor out $e^{-2mv^0\omega}$ from $e^{\frac{\mathcal{R}(M k, M^2 \omega)}{M^2}}$ since it is a constant ($v^0 = \tau^0 = 1$ in adapted coordinates). Following this, the $\omega$ integral can be easily evaluated
\begin{equation}
\int \limits_{0}^{\infty} \frac{d\omega}{2 \pi} e^{-2m\omega} = \frac{1}{4 \pi m} \, .
\label{an.om}
\end{equation}
Concerning the $k$ integral, we need to use the BCH expansion to factor out $e^{-k^2}$ from $e^{\frac{\mathcal{R}(M k, M^2 \omega)}{M^2}}$. By labelling $A= -k^2$ and $B$ as the $M$ dependent terms of $\frac{\mathcal{R}(M k, M^2 \omega)}{M^2}$ , we can write
\begin{equation}
e^{A+B} = e^Ae^E \, ,
\label{an.fact}
\end{equation}
where $E$ is given by
\begin{align}
E &= B - \frac{\left[A,B\right]}{2} + \frac{\left[A,\left[A,B\right]\right]}{6} + \frac{\left[B,\left[A,B\right]\right]}{12} -  \frac{\left[A,\left[B,\left[A,B\right]\right]\right]}{24} - \frac{\left[A,\left[A,\left[A,B\right]\right]\right]}{24} \notag\\
&\qquad + \frac{\left[A\left[A,\left[A,\left[A,B\right]\right]\right]\right]}{120} + \frac{\left[A\left[A,\left[B,\left[A,B\right]\right]\right]\right]}{120} - \frac{\left[A\left[B,\left[B,\left[A,B\right]\right]\right]\right]}{240}  \notag\\
&\qquad  + \frac{\left[B\left[A,\left[B,\left[A,B\right]\right]\right]\right]}{180} - \frac{\left[B\left[B,\left[B,\left[A,B\right]\right]\right]\right]}{720} + \frac{\left[B\left[A,\left[A,\left[A,B\right]\right]\right]\right]}{240} + \cdots \, .
\label{an.bch}
\end{align}
The ellipsis in Eq.~(\ref{an.bch}) refers to the fifth order onward terms of the BCH expansion. The commutators in Eq.~(\ref{an.bch}) contain all contributions up to $M^{-4}$ resulting from the BCH expansion, whose expressions have been provided in Eq.~(\ref{app3.bch}).  From Eq.~(\ref{app3.bch}) we see that all terms with even powers of $M^{-1}$ contain an even number of $k$'s and likewise all terms with odd powers of $M^{-1}$ contain an odd number of $k$'s. This property will hold to all orders in the BCH expansion.

Since $E$ contains $M^{-1}$ terms, we expand Eq.~(\ref{an.fact}) up to fourth order 
\begin{equation}
e^{A + B} = e^A \left( 1 + E + \frac{E^2}{2} + \frac{E^3}{3!} + \frac{E^4}{4!}\right) + \mathcal{O}(E^5)\, .\\
\label{an.exp}
\end{equation}
Eq.~(\ref{an.exp}) now contains all terms up to $M^{-4}$ which can contribute to the anomaly. We can now ignore all terms with free derivatives, as they cannot contribute to the anomaly. It will also be useful to separate those terms which do contain derivatives acting on $h^{-\frac{1}{4}}$ from those that do not. We thus write Eq.~(\ref{an.exp}) as 
\begin{align}
e^{A + B} &= e^A \left( 1 + E + \frac{E^2}{2} + \frac{E^3}{3!} + \frac{E^4}{4!}\right) + \mathcal{O}(E^5)\notag \\
&\approx e^A \left(1 + \frac{\mathcal{B}_1}{M} + \frac{\mathcal{B}_2}{M^2} + \frac{\mathcal{B}_3}{M^3} + \frac{\mathcal{B}_4}{M^4} + \mathcal{H}(h^{-\frac{1}{4}}) +\mathcal{O}(M^{-5})\right) \, . 
\label{an.exp2}
\end{align}
The $\approx$ symbol in Eq.~(\ref{an.exp2}) indicates that we have dropped all terms with free derivatives. $\mathcal{H}(h^{-\frac{1}{4}})$ contains all terms with $\partial(h^{-\frac{1}{4}})$, while the $\mathcal{B}_i$ terms represent the order $M^{-i}$ contributions which do not contain $\partial(h^{-\frac{1}{4}})$ . With Eq.~(\ref{an.exp2}), we have the following expression 
\begin{equation}
e^{\frac{\mathcal{R}(M k, M^2 \omega)}{M^2}} = e^{-2m\omega} e^{-k^2} \left(1 + \frac{\mathcal{B}_1}{M} + \frac{\mathcal{B}_2}{M^2} + \frac{\mathcal{B}_3}{M^3} + \frac{\mathcal{B}_4}{M^4} + \mathcal{H}(h^{-\frac{1}{4}}) \right) \, ,
\label{an.ok}
\end{equation}
which will be needed to evaluate the integrals. Upon substituting Eq.~(\ref{an.ok}) and Eq.~(\ref{an.om}) in Eq.~(\ref{an.ltor}), we get
\begin{align}
&\lim_{M \to \infty}\text{Tr}\Lambda(x) e^{\frac{\mathcal{R}}{M^2}} \notag\\&
= \lim_{M \to \infty} M^4 \frac{1}{4 \pi m} \int  \frac{d^2 k}{(2\pi)^2} \Lambda(x) e^{-k^2} \left(1 + \frac{\mathcal{B}_1}{M} + \frac{\mathcal{B}_2}{M^2} + \frac{\mathcal{B}_3}{M^3} + \frac{\mathcal{B}_4}{M^4} + \mathcal{H}(h^{-\frac{1}{4}}) \right)
\label{an.main}
\end{align}
Eq.~(\ref{an.main}) can now be evaluated via the following Gaussian integrals
\begin{align}
\int  d^2 k  \, e^{-k^2} =  \sqrt{h} \pi \, , &\, \int  d^2 k  \, e^{-k^2} k_i k_j = \frac{1}{2} \sqrt{h} \pi h_{ij} \notag\\
 \int  d^2 k  \, e^{-k^2} k_i k_jk_mk_n &= \frac{1}{4} \sqrt{h} \pi \left(h_{ij}h_{mn} + h_{im}h_{nj} + h_{in}h_{mj} \right) \notag\\
\int  d^2 k  \, e^{-k^2} \, k_i k_j \cdots k_{2n-1} k_{2n} &= \frac{1}{2^n} \sqrt{h} \pi \left( (2n-1)!!\, \, \text{permutations of} \, \, h_{ij} \cdots h_{2n-1 \,  2n} \right)\, .
\label{an.symint}
\end{align}
The $k$ integrals vanish under symmetric integration whenever there are an odd number of $k$'s in the integrand. Thus $\mathcal{B}_1$ and $\mathcal{B}_3$ vanish under symmetric integration. $\mathcal{H}(h^{-\frac{1}{4}})$ also vanishes following symmetric integration. This result could have been anticipated from the cyclicity of trace \footnote{The argument involving the cyclicity of trace works in the present case as the Jacobian did not involve any free derivatives ($J= \Lambda$).}. The integral 
\begin{equation}
\int  d^2 k e^{-k^2} \left(1 + \frac{\mathcal{B}_2}{M^2}\right) \, ,
\end{equation}  
is non-vanishing. These terms would be eliminated by regularization in a one-loop calculation and do not contribute in the final expression for the anomaly. For example, within the Pauli-Villars scheme one can include additional copies of the PV fields with coefficients chosen so as to cancel out these $M$ dependent terms. Thus these terms can be ignored as well. Since the integral of $\mathcal{B}_2$ is somewhat instructive, we have provided the terms contained in its integrand in Eq.~(\ref{app3.b2}), using which we have the following result
\begin{equation}
\int  d^2 k e^{-k^2}\frac{\mathcal{B}_2}{M^2} = \sqrt{h}\pi \left(\frac{1}{6} R_{ij} h^{ij} + 2 m^2 \phi \right) \,.
\label{an.b2}
\end{equation} 
The only contribution to the anomaly comes from the term $\mathcal{B}_4$ and Eq.~(\ref{an.main}) reduces to
\begin{equation}
\lim_{M \to \infty}\text{Tr}\Lambda(x) e^{\frac{\mathcal{R}}{M^2}} =  \frac{1}{4 \pi m} \int  \frac{d^2 k}{(2\pi)^2} \Lambda(x) e^{-k^2} \mathcal{B}_4
\label{an.red}
\end{equation}
The individual terms contained in $\mathcal{B}_4$ have been provided in Eq.~(\ref{app3.b4}), and the resulting $k$ integral works out to give
\begin{align}
\int d^2k  e^{-k^2} \mathcal{B}_4 &= \sqrt{h} \pi \left(\frac{1}{180} (R_{ijmn}R^{ijmn} -  R_{ij}R^{ij} + \Box R_{ij}h^{ij})  \right. \notag\\
&\left. \qquad \qquad  \qquad \qquad +  2 m^4 \phi^2 + \frac{m^2}{3} (\phi R_{ij}h^{ij} + R_{00}v^0 v^0)\right) \, .
\label{an.b4}
\end{align}
Substituting Eq.~(\ref{an.b4}) in Eq.~(\ref{an.red}), we get the following expression for the candidate anomaly,
\begin{align}
\lim_{M \to \infty}\text{Tr}\Lambda(x) e^{\frac{\mathcal{R}}{M^2}} &= \frac{\sqrt{h}\Lambda(x)}{m (4\pi)^2}\left(\frac{1}{180} (R_{ijmn}R^{ijmn} -  R_{ij}R^{ij} + \Box R_{ij}h^{ij})  \right. \notag\\
&\left. \qquad \qquad  \qquad \qquad +  2 m^4 \phi^2 + \frac{m^2}{3} (\phi R_{ij}h^{ij} + R_{00}v^0 v^0)\right)
\label{an.ca}
\end{align}
While the calculation leading to this result is considerably involved, we note the following points related to the derivation and the above result. The term $R_{00}v^0v^0$ results due to both the single derivative operator $\partial_t$ and $im\mathcal{C}$ contained in Eq.~(\ref{an.regm}), following the BCH expansion. If $A_{\mu}$ were absent in our derivation, then so too would all the terms in the second line of Eq.~(\ref{an.ca}), thereby providing only the curvature squared results already noted in the literature. The choice of $\tau_{\mu} = (1,0,0,0)$ and the absence of $h^{0 \mu}$ in adapted coordinates affects the expressions of $\mathcal{C}$, the Ricci and Riemann tensors, as well as the final result. The absence of terms $v^i$ and $\bar{\phi}$ in the final answer is thus a coordinate artifact which reflects our choice of time for the hypersurface. Remarkably, all imaginary terms cancel out in the calculation leading to Eq.~(\ref{an.ca}). The absence of imaginary terms as well as the split into ``temporal" and ``spatial" parts in the expression may also be noted in Eq.~(\ref{an.b2}).

The curvature squared terms of Eq.~(\ref{an.ca}) can be further simplified. We first note that a local counterterm involving $(R_{ij}h^{ij})^2$ can be included in the effective action to eliminate the term $\Box R_{ij}h^{ij}$. Hence this is not part of the final anomaly result. Further, since $(R_{ijmn}R^{ijmn} -  R_{ij}R^{ij})$ is constructed out of the $2d$ spatial metric on a NC background which satisfies the Frobenius condition, we can use $R_{ijmn} = \frac{1}{2}(R_{ij}h^{ij})\left(h_{im}h_{jn} - h_{in}h_{jm}\right)$ to write  
\begin{equation}
R_{ijmn}R^{ijmn} -  R_{ij}R^{ij} = \frac{1}{2}\left(R_{ij}h^{ij}\right)^2 \, .
\label{an.r2d} 
\end{equation}
Thus using Eq.~(\ref{an.ca}) and Eq.~(\ref{an.an1}) we can write the following covariant result
\begin{equation}
\left \langle 2 \widetilde{T}^{0}_{\phantom{\mu} 0} + \widetilde{T}^{i}_{\phantom{\mu} i} \right \rangle = \frac{1}{m (4\pi)^2}\left(\frac{1}{360} (R_{\mu \nu}h^{\mu \nu})^2  +  2 m^4 \psi^2 + \frac{m^2}{3} (\psi R_{\mu \nu}h^{\mu \nu} + R_{\mu \nu}v^{\mu} v^{\nu})\right)
\label{an.tran}
\end{equation}
In going from Eq.~(\ref{an.ca}) to Eq.~(\ref{an.tran}) we have accounted for the presence of $h^{0 \mu}$ terms which should have been present in the expressions of $v^0$ and $\phi$ for coodinate choices other than adapted coordinates. This covariant result has been inferred from Eq.~(\ref{an.ca}) by noting the Milne invariance of the regulator we considered in Eq.~(\ref{an.regm}) , as well as the absence of `Milne gravitational anomalies' \footnote{As the field $\Phi$ does not transform under Milne transformations, there can be no corresponding anomaly even though the action itself is Milne invariant}. While in principle \emph{any} $h^{0\mu}$ contribution of Eq.~(\ref{an.ca}) could have been involved in the final answer, only one specific choice leads to the Milne invariant result of Eq.~(\ref{an.tran}). 

On the other hand, it may be noted that Eq.~(\ref{an.tran}) violates $U(1)$ invariance. We have been unable to find counterterms which would help eliminate the $\psi R_{\mu \nu}h^{\mu \nu}$ and $\psi^2$ terms of Eq.~(\ref{an.tran}) and they do appear to comprise the true anomaly.  A similar situation arises in relativistic systems which involve gauge and gravitational anomalies. A characteristic example arises in the 4 dimensional mixed gravitational anomaly which involves both $U(1)$ and diffeomorphism anomalies, where the latter violates $U(1)$ invariance. However, one can find a counterterm to make the gauge current anomaly free, which in turn leads to the diffeomorphism anomaly being $U(1)$ invariant \cite{Iqbal:2015vka}. We believe a situation similar to this would arise for the NC background. The key difference with the relativistic case is that the anomalous current $\langle J^{\mu} \rangle$ is also a gravitational anomaly due to its presence in the connection. 

To conclude, we point out some further generalities which may be deduced from our calculation. We note that the trace anomaly can only arise in odd dimensions. Since $z=2$ and all BCH expansion terms involve an even (odd) number of $k$'s for terms with an even (odd) power of $M^{-1}$, the anomalies can only occur when there are an even number of spatial dimensions. Thus NC trace anomalies always arise in odd spacetime dimensions.

While our result concerned NC backgrounds without torsion, which allowed us to use Eq.~(\ref{an.r2d}), in general we would have instead
\begin{equation}
R_{ijmn}R^{ijmn} -  R_{ij}R^{ij} = \frac{1}{2}(-\bar{E}_4 + 3 \bar{C}^2) \, ,
\end{equation} 
where $E_4$ and $C^2$ represent the four dimensional Euler density and the square of the Weyl tensor respectively as follows
\begin{align}
E_4 &= R_{\mu \nu \rho \sigma}R^{\mu \nu \rho \sigma} - 4 R_{\mu \nu}R^{\mu \nu} + R^2 \, ,\notag\\
C^2 &= R_{\mu \nu \rho \sigma}R^{\mu \nu \rho \sigma} - 2 R_{\mu \nu}R^{\mu \nu} + \frac{1}{3}R^2 \, ,
\end{align}
while the overbar implies that these tensors are contracted only with the (two dimensional) spatial metric $h^{\alpha \beta}$. The general result, following Eq.~(\ref{an.tran}), will then be modified to 
\begin{align}
\left \langle 2 \widetilde{T}^{0}_{\phantom{\mu} 0} + \widetilde{T}^{i}_{\phantom{\mu} i} \right \rangle &= \frac{1}{m (4\pi)^2}\left(\frac{1}{360} \left(-\bar{E}_4 + 3 \bar{C}^2\right)  +  2 m^4 \psi^2 + \frac{m^2}{3} (\psi R_{\mu \nu}h^{\mu \nu} + R_{\mu \nu}v^{\mu} v^{\nu})\right)\notag\\& \qquad\qquad + \text{additional terms} \, .
\label{an.tran2}
\end{align}
This result, apart from the $\tau^{\mu}$ and $A^{\mu}$ dependent terms is in agreement with the results provided in \cite{Jensen:2014hqa,Auzzi:2015fgg}. 
The coefficients of the curvature squared terms are in addition identical to those derived using the heat kernel approach of \cite{Auzzi:2016lxb}.\\

\subsection{The diffeomorphism anomaly}
The diffeomorphism anomaly can be computed from Eq.~(\ref{sf.act}) using the procedure of the previous subsection. The fundamental fields are now $\Phi$ and $\Phi^*$ with the following action
\begin{equation}
S = \int dt d^2x {\Phi}^* \sqrt{h} \mathcal{D}  \Phi  \notag\\
\label{an.act2}
\end{equation}
The path integral in this case is given by,
\begin{equation}
Z = \int \mathcal{D}\Phi \mathcal{D}\Phi^* e^{i S\left[\Phi, \Phi^*, \tau^{\mu}, h^{\mu\nu}, A_{\mu}\right]} \, 
\label{an.pi2}
\end{equation}
Using Eq.~(\ref{sf.diff}), the invariance of Eq.~(\ref{an.pi2}) under $\delta\Phi=\pounds_{\xi} \Phi$ and $\delta \Phi^*=\pounds_{\xi} \Phi^*$ results in the following anomalous Ward identity
\begin{equation}
 \left\langle- \sqrt{h}\xi^{\mu}\left(\nabla_{\nu}T^{\nu}{}_{\mu} + J^{\nu}\nabla_{[\mu}A_{\nu]} - \frac{1}{2} R_{\nu}\nabla_{\mu}\tau^{\nu}\right)\right\rangle_{\Phi \Phi^*} = \left\langle\text{Tr}J\right\rangle_{\Phi \Phi^*} \, .
\label{an.regdt}
\end{equation}
From Eq.~(\ref{an.pact}) we have $\mathcal{R} = \mathcal{D}$, which ensures that it is symmetric\footnote{By symmetric we mean that $\int \Phi^* \mathcal{D} \Phi = \int \Phi \bar{\mathcal{D}} \Phi^*$}. Here $\mathbf{T} = 2 \sqrt{h}$ and hence from Eq.~(\ref{an.jaco}) the Jacobian to consider is
\begin{align}
J &= \xi^{\mu}\partial_{\mu} + \frac{1}{2\sqrt{h}}\pounds_{\xi}\sqrt{h}  \notag\\
&= \xi^{\mu}\partial_{\mu} + \frac{1}{2\sqrt{h}} \xi^{\mu}\partial_{\mu}\sqrt{h} +  \partial_{\mu}\xi^{\mu} \, .
\label{an.jac2}
\end{align}
Thus the regulated trace takes the following form
\begin{align}
&\lim_{M \to \infty}\text{Tr} J e^{\frac{\mathcal{R}}{M^2}}  
\notag\\&= \lim_{M \to \infty} \int \frac{d\omega}{2 \pi} \int \frac{d^2k}{(2 \pi)^2} e^{-i \omega t} e^{ikx}\left[\left(\xi^{\mu}\partial_{\mu} + \frac{\xi^{\mu}\partial_{\mu}\sqrt{h}}{2\sqrt{h}}  +  \partial_{\mu}\xi^{\mu}\right) e^{\frac{\mathcal{R}}{M^2}} \right] e^{i \omega t} e^{- ikx} \, .
\label{an.dtr}
\end{align}
Evaluating this expression would formally result in considering an expansion up to $M^{-6}$. In taking the plane wave $(e^{i \omega t} e^{- ikx})$ to the left, it gets acted upon by both the Jacobian and the regulator. The action of the Jacobian on $e^{i \omega t}$ now produces the term $i \xi^0 \omega$ . By rescaling $\omega \to M^2 \omega$, we end up with a factor of $M^6$ outside the above integral, requiring a BCH expansion up to $M^{-6}$ for determining the anomaly.

~~However, having chosen a symmetric regulator we can avoid this cumbersome calculation by noting the following identity which holds for any symmetric regulator $\widetilde{\mathcal{R}}$ \cite{Hatsuda:1989qy}
\begin{equation}
\text{Tr}\left(\xi^{\mu}\partial_{\mu} + \frac{1}{2}\partial_{\mu}\xi^{\mu} \right)e^{\widetilde{\mathcal{R}}} = 0 \, .
\label{an.symid}
\end{equation}
Using the expressions for $\Gamma^i_{\mu i} = \frac{1}{\sqrt{h}}\partial_{\mu}\sqrt{h}$ and $\Gamma^{0}_{\mu \nu} = 0$ (in adapted coordinates) and Eq.~(\ref{an.symid}), we can simplify Eq.~(\ref{an.dtr}) to
\begin{equation}
\lim_{M \to \infty}\text{Tr} J e^{\frac{\mathcal{R}}{M^2}}  = \lim_{M \to \infty} \int  \frac{d\omega}{2 \pi} \int \frac{d^2k}{(2 \pi)^2} e^{-i \omega t} e^{ikx}\left[\frac{1}{2}\left(\nabla_{\mu}\xi^{\mu}\right) e^{\frac{\mathcal{R}}{M^2}} \right] e^{i \omega t} e^{- ikx} \, .
\label{an.dtr2}
\end{equation}
Hence we don't have to deal with any free derivatives due to the Jacobian. Moving the plane wave past the regulator and rescaling $k \to M k$ and $\omega \to M^2 \omega$ results in
\begin{equation}
\lim_{M \to \infty}\text{Tr} J e^{\frac{\mathcal{R}}{M^2}} = \lim_{M \to \infty} M^4 \int \frac{d\omega}{2 \pi} \int  \frac{d^2 k}{(2\pi)^2} \frac{1}{2}\left(\nabla_{\mu}\xi^{\mu}\right) e^{\frac{\mathcal{R}(M k, M^2 \omega)}{M^2}} \, 
\label{an.ltor2}
\end{equation}

We now need to factor out $e^{-k^2}$ and $e^{-2m\omega}$ from $e^{\frac{\mathcal{R}(M k, M^2 \omega)}{M^2}}$ using the BCH expansion, as in the previous section, up to $M^{-4}$ terms. Since the regulator of this subsection differs from that of the previous one only by $\partial(h^{-\frac{1}{4}})$ terms, the following factored expression is easily determined from Eq.~(\ref{an.ok})
\begin{equation}
e^{\frac{\mathcal{R}(M k, M^2 \omega)}{M^2}} = e^{-2m\omega} e^{-k^2} \left(1 + \frac{\mathcal{B}_1}{M} + \frac{\mathcal{B}_2}{M^2} + \frac{\mathcal{B}_3}{M^3} + \frac{\mathcal{B}_4}{M^4} \right) \, .
\label{an.ok2}
\end{equation}
Only the $\mathcal{B}_4$ term contributes to the anomaly, and we have the following expression for the candidate anomaly 
\begin{align}
&\lim_{M \to \infty}\text{Tr} J e^{\frac{\mathcal{R}}{M^2}}= \frac{1}{2}\left(\nabla_{\mu}\xi^{\mu}\right)\int \frac{d\omega}{2 \pi} e^{-2m\omega} \int \frac{d^2k}{(2\pi)^2}  e^{-k^2} \mathcal{B}_4 \notag\\
&= \frac{\sqrt{h}\left(\nabla_{\mu}\xi^{\mu}\right)}{(4 \pi)^2 m}\left(\frac{1}{360} \left((R_{ij}h^{ij})^2 + \Box(R_{ij}h^{ij})\right) +  2 m^4 \phi^2 + \frac{m^2}{3} (\phi R_{ij}h^{ij} + R_{00}v^0 v^0)\right) 
%
\label{an.b42}
\end{align}
where we have simplified the curvature squared expression by making use of Eq.~(\ref{an.r2d}) . The terms from Eq.~(\ref{an.b42}) which contribute to the anomaly must satisfy the same criteria as in the case for the trace anomaly. Adopting the covariant notation as in the case of the trace anomaly, the result for the diffeomorphism anomaly in this case is
\begin{align}
\text{An}_{\xi} &= -\frac{\sqrt{h}}{m (4\pi)^2}\xi^{\mu}\nabla_{\mu}\left(\frac{1}{720} (R_{\mu \nu}h^{\mu \nu})^2  +   m^4 \psi^2 + \frac{m^2}{6} (\psi R_{\mu \nu}h^{\mu \nu} + R_{\mu \nu}v^{\mu} v^{\nu})\right) \, .
\label{an.caf2}
\end{align}
Eq.~(\ref{an.caf2}) and Eq.~(\ref{an.regdt}) now provide the following expression for the diffeomorphism anomaly
\begin{align}
& \left\langle \nabla_{\nu}T^{\nu}{}_{\mu} + J^{\nu}\nabla_{[\mu}A_{\nu]} - \frac{1}{2} R_{\nu}\nabla_{\mu}\tau^{\nu}\right\rangle \notag\\
& =  \nabla_{\mu}\left(\frac{1}{720 (4\pi)^2 m} (R_{\alpha \beta}h^{\alpha \beta})^2  +\frac{m^3}{16\pi^2} \psi^2+ \frac{m}{96 \pi^2} (\psi R_{\mu \nu}h^{\mu \nu} + R_{\mu \nu}v^{\mu} v^{\nu})\right) 
\label{an.fin2}
\end{align}
We emphasize that all currents occurring on the left hand side of Eq.~(\ref{an.fin2}) correspond to the gravitational fields of the NC background. We note that most of the previous results for the trace anomaly (based on DLCQ) indicate a one-to-one correspondence of the $2+1$ dimensional result of the NC background with $3+1$ dimensional result of relativistic backgrounds. Were this to actually be true for all gravitational anomalies, one would in fact naively expect there to be no diffeomorphism anomaly for the Schr\"odinger field in $2+1$ dimensions. In deriving this result, we have demonstrated that this is not the case. The presence of a diffeomorphism anomaly allows for several consequences in condensed matter systems with boundaries. In particular we note that this could be relevant in providing the entanglement entropy of Quantum Hall systems on curved backgrounds with boundaries \cite{Iqbal:2015vka,Hughes:2015ora}, where the Schr\"odinger field is present in the low energy effective action.
%
\section{A c-theorem condition} \label{cthm}
The coefficients of the trace anomaly are closely related to the renormalization group (RG) flow of a given theory.
By applying the Wess-Zumino (WZ) consistency condition on the quantum effective action one can relate the anomaly coefficients with the beta functions of the theory. Our treatment in this section will follow \cite{Jack:1990eb} where the consistency conditions for 2d and 4d relativistic CFTs were addressed. An investigation of the local RG flow due to the curvature squared terms of Eq.~(\ref{an.tran2}) was considered in \cite{Auzzi:2016lrq}. Here we confine our attention to the $U(1)$ invariant term $R_{\mu \nu} \tau^{\mu}\tau^{\nu}$ contained in Eq.~(\ref{an.tran}). Our goal in this section will be to demonstrate that this term satisfies a c-theorem condition analogous to that of 2d CFTs. To begin with, let us consider the following renormalized partition function in the presence of sources,
\begin{equation}
Z\left[\mathcal{J}\right] = e^{i \mathcal{W}\left[\mathcal{J}\right]} = \int \mathcal{D}\widetilde{\Phi} \mathcal{D}\widetilde{\Phi}^* e^{i S\left[\widetilde{\Phi},\widetilde{\Phi}^*, \mathcal{J}\right]} \, ,
\end{equation}
where $\mathcal{W}$ is the quantum effective action, which generates connected correlators associated with renormalized composite operators, and $\mathcal{J}$ denotes all the sources. Here we will assume that $\mathcal{J}$ involves the independent background fields of the NC backround  ($h^{\mu\nu} \, , \tau^{\mu} \,, \tau_{\mu}$ and  $A_{\mu}$) and dimensionless coefficients $g^I$ associated with certain marginal operator insertions $\mathcal{O}_I$ \footnote{In general $\mathcal{J}$ also involves $m^a$ associated with relevant operators $\mathcal{O}_a$, and vector sources $\mathcal{A}_{\mu}$ associated with certain currents $\mathscr{J}^{\mu}$ which the theory might possess}. To investigate RG flows we first introduce the RG parameter $\mu$. We can now define the RG time function $t =  \text{ln} \left(\frac{\mu}{\mu_0}\right)$, where $\mu_0$ is some arbitrary reference scale and the beta functions $\beta^I = \frac{\partial g^I}{\partial t}$ correspond to the dimensionless parameters $g^I$. The flow is generated by $\mathscr{D} = \mu \frac{\partial}{\partial \mu} + \beta^I \partial_I$, where we have further defined $\partial_I = \frac{\partial }{\partial g^I}$. In flat spacetime $\mathcal{W}$ satisfies the flow equation
\begin{equation}
\mathscr{D}\mathcal{W} = 0 \, ,
\end{equation}
which is nothing but the Callan-Symanzik equation. The local RG concerns itself with the renormalizability of composite operators on curved backgrounds and hence the couplings are now functions of spacetime ($g^I = g^I(x,t)$). The local Callan-Symanzik equation under Weyl transformations is given by
\begin{equation}
\left(\Delta^W_{\Lambda} - \Delta^{\beta}_{\Lambda}\right)\mathcal{W} = \int \limits_{\mathcal{V}} dv B_{\Lambda} \, ,
\label{rg.lcs}
\end{equation}
where $\Lambda$ is the local parameter involved in Weyl transformations, $\int \limits_{\mathcal{V}} dv$ is the integral involving the NC covariant volume element in ($2+1$) dimensions and $B_{\Lambda}$ is a local anomaly density involving derivatives of the NC fields and $g^I$. The variations $\Delta^W_{\Lambda}$ and $\Delta^{\beta}_{\Lambda}$ are defined as
\begin{align}
\Delta^W_{\Lambda} &= \int \limits_{\mathcal{V}} dv \left[ 2 \Lambda h^{\mu \nu} \frac{\delta}{\delta h^{\mu \nu}} + 2 \Lambda \tau^{\mu} \frac{\delta}{\delta \tau^{\mu}} \right] \notag \\
\Delta^{\beta}_{\Lambda} &= \int \limits_{\mathcal{V}} dv  \Lambda \beta^I \frac{\delta}{\delta g^I} \, 
\end{align}
Eq.~(\ref{rg.lcs}) reveals that at the critical point, where $\beta^I = 0$, $B_{\Lambda}$ is simply the trace anomaly. Away from the critical point, we have additional dimension $4$ terms involving the derivatives $g^I$. We can thus write Eq.~(\ref{rg.lcs}) in the following way
\begin{equation} 
\left(\Delta^W_{\Lambda} - \Delta^{\beta}_{\Lambda}\right)\mathcal{W} = \int \limits_{\mathcal{V}} dv \tau^{\mu} \tau^{\nu}\left[\Lambda \left(\frac{1}{2}\beta^{\Phi} R_{\mu \nu} - \frac{1}{2}\chi_{IJ}\partial_{\mu}g^I \partial_{\nu}g^J \right) - \left(\partial_{\mu}\Lambda\right)\omega_I \partial_{\nu} g^I + \cdots \right] \, ,
\label{rg.cthm}
\end{equation}
where $\beta^{\Phi} \, , \chi_{IJ}$ and $\omega_I$ all depend on the coupling parameter $g^I$. The dots in Eq.~(\ref{rg.cthm}) indicate all anomaly terms of Eq.~(\ref{an.tran}) other than $R_{\mu \nu}\tau^{\mu}\tau^{\nu}$, as well as additional terms of dimension $4$. These terms have been ignored since they will not be required in the following discussion. As before, we assume that the NC background satisfies the Frobenius condition. Since Weyl transformations are Abelian, they satisfy the WZ consistency condition
\begin{equation}
\left[\Delta^W_{\Lambda} - \Delta^{\beta}_{\Lambda}\, , \, \Delta^W_{\Lambda'} - \Delta^{\beta}_{\Lambda'} \right]\mathcal{W} = 0 \, ,  
\label{rg.wz}
\end{equation}
Using Eq.~(\ref{rg.cthm}), Eq.~(\ref{rg.wz}) gives the following expression
\begin{equation}
\left[\Delta^W_{\Lambda} - \Delta^{\beta}_{\Lambda}\, , \, \Delta^W_{\Lambda'} - \Delta^{\beta}_{\Lambda'} \right]\mathcal{W} = \int \limits_{\mathcal{V}} dv \tau^{\nu}\left(\Lambda \partial_{\nu} \Lambda' - \Lambda' \partial_{\nu} \Lambda\right)\tau^{\mu}V_{\mu}  = 0\, .
\label{rg.null}
\end{equation}
where
\begin{equation}
V_{\mu} = \partial_{\mu}\beta^{\Phi} - \left( \chi_{IJ}\beta^I - \beta^I \partial_I \omega_J - \omega_I \partial_J\beta^I \right) \partial_{\mu}g^J
\label{rg.main}
\end{equation}
Eq.~(\ref{rg.null}) is satisfied when $V_{\mu}$ vanishes. This implies
\begin{equation}
\partial_{J}\beta^{\Phi} = \chi_{IJ}\beta^I - \beta^I \partial_I \omega_J - \omega_I \partial_J\beta^I \, .
\label{rg.eq}
\end{equation}
We now define the new function $\widetilde{\beta}^{\Phi} = \beta^{\Phi} + \omega_I\beta^I$, with which Eq.~(\ref{rg.eq}) becomes
\begin{equation}
\partial_{J}\widetilde{\beta}^{\Phi} = \chi_{IJ}\beta^I + \beta^I \left(\partial_J \omega_I -\partial_I \omega_J\right) \, ,
\end{equation}
Contracting this equation with $\beta^J$ now leads to the following result
\begin{equation}
\frac{\partial \widetilde{\beta}^{\Phi}}{\partial t} = \chi_{IJ} \beta^I \beta^J
\label{rg.rgf}
\end{equation}
This is a c-theorem condition satisfied by the coefficient of $R_{\mu \nu}\tau^{\mu} \tau^{\nu}$ on NC backgrounds with the Frobenius condition, which is analogous to the relation satisfied in 2d CFTs.  At this point the proof of the c-theorem follows by establishing that the `metric' $\chi_{IJ}$ is positive definite. 
 In 2d CFTs, it can be shown that $\chi_{IJ}$ is essentially equivalent to `Zamolodchikov's metric' $G_{IJ} = (x^2)^2\langle\left[\mathcal{O}_I(x)\right] \left[\mathcal{O}_J(0)\right]\rangle$, which further identifies $\widetilde{\beta}^{\Phi}$ with Zamolodchikov's c-function $C$ \cite{Jack:1990eb}. Here the situation is not so straightforward since the marginal operators and the correlation functions they define differ from those of 2d CFTs. Our analysis would also be incomplete without all the terms of Eq.~(\ref{an.tran}) and their consistency conditions. As these considerations lies outside the scope of the present work, we will address them in the future.
\section{Discussion} \label{sec.con}
We have derived the trace and diffeomorphism anomalies of the Schr\"odinger field minimally coupled to the NC background in $2+1$ dimensions following Fujikawa's approach. In doing so, we determined that the modified coupling necessary to render the Schr\"odinger action invariant under Milne boosts was necessary in the derivation of a curvature term similar to the relativistic anomaly of one less spacetime dimension. In the language of \cite{Deser:1993yx} this is a Type A anomaly, which distinguishes the result from those of Lifshitz backgrounds with the Frobenius condition \cite{Arav:2016xjc}. The expression for the trace anomaly in addition contains curvature squared terms, contracted with the spatial metric of the background. As demonstrated, without the Frobenius condition these curvature squared terms always provide both Type A and Type B anomaly terms.

Collectively, the anomaly we found may be considered as a general expression built out of curvature invariants of mass dimension $4$ which contain ``spatial" and ``temporal" contributions. We conjecture that in $d+1$ spacetime dimensions, where $d = 2n \, ; n=1,2,\cdots$, the result will contain an expression of the $d+2$ dimensional relativistic anomaly, contracted with the spatial metric, along with a term of the form of the $d$ dimensional relativistic anomaly, contracted in general with both spatial and temporal metrics, such that all terms are individually of mass dimension $d+2$.  We believe that the result of Eq.~(\ref{an.tran2}) should follow from a heat kernel approach which fully accounts for the field $A_{\mu}$ and its variations. Our final expression for the trace and diffeomorphism anomalies also contained specific $U(1)$ violating terms. These terms need to be understood in the context of the $U(1)$ anomaly which was not derived here. Due to the presence of the $A_{\mu}$ field in the connection, its derivation will involve a regulator quite different from those considered in the relativistic case. 

The coefficients of the trace anomaly will have interesting implication for field theories on the NC background. This is evident from the $m$ dependence in Eq.~(\ref{an.tran}) which indicates that the curvature and curvature squared terms dominate in different regimes. Specifically, for $0<m<1$ the curvature squared contribution dominates, while for $m > 1$ we have the dominant contribution from the $R_{\mu \nu} \tau^{\mu}\tau^{\nu}$ term. We also note that the coefficient of the $R_{\mu \nu} \tau^{\mu}\tau^{\nu}$ term, apart from a factor of $\frac{m}{\pi}$, is precisely one half of that of the relativistic trace anomaly in $1+1$ dimensions. One way in which we can appreciate the physical implications of these observations is through the local RG flow. In the previous section, we demonstrated that the coefficient of the $R_{\mu\nu}\tau^{\mu}\tau^{\nu}$ term satisfies an expression analogous to the c-theorem of 2d CFTs. We however did not identify the metric of parameter space ($\chi_{IJ}$) occuring in this expression with a manifestly positive definite quantity related to the correlation functions of the fields. These correlators can be derived from those of the Klein Gordon field in the limit of $\frac{m}{E} >>1$. One therefore expects that massive deformations about the fixed point of the Schr\"odinger theory might also be relevant to the RG flow. While the coefficient of $(R_{\mu \nu}h^{\mu \nu})^2$ will have to vanish at the fixed point in order to satisfy the WZ consistency conditions, its behaviour away from the fixed point might be influenced by such deformations. We look forward to provide a detailed analysis of this point as well as the RG flow on NC backgrounds in future work.

NC gravitational anomalies will also be relevant for certain systems with boundaries. As the AdS/CFT correspondence is expected to hold in the NR limit~\cite{Balasubramanian:2010uw,Guica:2010sw,Kim:2012nb,Detournay:2012dz,Andrade:2014kba,Hartong:2014pma}, the bulk anomalies in $2+1$ dimensions will impose certain constraints on the nature of the dual field theory at the boundary. It will be intersting to consider possible differences with Lifshitz holography~\cite{Taylor:2015glc} as the anomalies found in this work differ from those of Lifshitz backgrounds. We also know that the low energy effective action for Quantum Hall systems involves the Schr\"odinger field coupled to, in general, a background gravitational field in $d+1$ dimensions; where $d=2n$. Anomalies play a crucial role in Hall phenomenology~\cite{Stone:2012cx,Stone:2012ud,Gromov:2014gta,Gromov:2014vla,Gromov:2014dfa,Can:2014awa,Can:2014ota,Iqbal:2015vka,Hughes:2015ora,Gromov:2015fda}, with the guiding principle in the presence of boundaries being that the bulk and boundary contributions collectively should be non-anomalous~\cite{Stone:2012ud}. In taking the NR limit for these systems, we have a bulk gravitational anomaly and no boundary anomaly. One can thus expect that the cancellation of anomalies in this case might manifest in certain surface effects.
\section{Acknowledgment}
We thank Roberto Auzzi, Giuseppe Nardelli and Sridip Pal for many useful comments and Michael Stone for stimulating discussions on anomalies and their applications.\\

\appendix
\section{Adapted coordinates for the NC background} \label{app.ac}
%
Relativistic gravitational anomalies using Fujikawa's approach can be calculated in a covariant notation in a local plane wave basis. In the NR case, we do need to distinguish between time and space in the plane waves as well as the regulator. We thus need to make use of a specific set of coordinates in our calculation. The adapted coordinates \cite{D} provides a representation of the NC structure.
Let Greek indices $\mu,\nu,\cdots$ denote spacetime coordinates, Latin indices $i,j,\cdots$ denote spatial coordinates and $0$ represent the coordinate for time. Then the NC system of equations for the metric can be realized through the following choice
\begin{align}
\tau_0 = 1 =  \tau^{0} \, , \quad  \tau_i = 0   \, ,\quad h^{0 \mu} = 0 \label{nc.tcoord}
\end{align}
Eq.~(\ref{nc.tcoord}) represents our choice of time. The normalization of $\tau_{\mu}$ Eq.~(\ref{nc.norm}) allows us to make the choice given in Eq.~(\ref{nc.tcoord}). Since $A_{\mu}$ is a gauge field, it is naturally left unspecified.
 Using Eq.~(\ref{nc.con}) and Eq.~(\ref{nc.tcoord}), we have the following non-vanishing components for the connection
\begin{align}
&\qquad \qquad \qquad \qquad \qquad \quad \Gamma^i_{j k} = \big \{^{\phantom{i} i}_{jk} \big \} \, ,\notag\\
\Gamma^i_{0 j} = & \frac{h^{ik}}{2}\left(\partial_j h_{k0}+ \partial_0 h_{kj} - \partial_k h_{0j} - \partial_kA_j + \partial_j A_k \right) \, , \quad \Gamma^i_{0 i} = \frac{h^{ik}}{2}  \partial_0h_{ik} \, , \notag\\
& \qquad \qquad \Gamma^i_{0 0} = \frac{h^{i k}}{2} \left(2\partial_0h_{k0} - \partial_k h_{00}\right) - h^{i k} \left(\partial_k \phi - \partial_0 A_k \right)   \, ,
\label{app.conn}
\end{align}
where $\big \{^{\phantom{i} i}_{jk} \big \}$ represents the ``Christoffel" component of the connection for the spatial metric (the second term of Eq.~(\ref{nc.con})). 
Notably, $h_{0 \mu}$ need not vanish in adapted coordinates and therefore $\tau^i$ can exist. Using Eq.~(\ref{nc.proj}) and Eq.~(\ref{nc.tcoord}) we find that $h_{\mu\nu}$ and $\tau^{\mu}$ satisfy the following relations
\begin{align}
h_{ij}\tau^{j} &= -h_{i0}, ~~\tau^{i} = -h^{ij}h_{j0} \, ,\notag\\
h_{00} &= -h_{0j}\tau^j = \tau^i h_{ij} \tau^j \, .
\end{align}
It can now be seen that the mass dimension of the connection components in Eq.~(\ref{app.conn}) are not the same. The first line of Eq.~(\ref{app.conn}) has mass dimension 1, the second line has mass dimension 2, while the last line has mass dimension 3. This reflects the $z=2$ invariance of the background. However, Ricci and Riemann tensor components have a uniform mass dimension as a consequence. For instance 
\begin{equation}
R_{00} = \Gamma^i_{00,i} - \Gamma^i_{0 i, 0} + \Gamma^i_{i j}\Gamma^j_{0 0} - \Gamma^i_{0 j}\Gamma^j_{0 i} \, ,
\label{app.r00}
\end{equation}
has mass dimension 4, while $R_{ij}$ has mass dimension $2$.
\section{Fujikawa's approach and Regulators} \label{app2.app2}
Here we review the background material needed for the calculation of anomalies provided in Sec.~\ref{anom}. Our arguments will be catered to address the gravitational anomalies considered in this paper. 

\subsection{Fujikawa's approach} \label{app2.fuji}
Anomalies can be understood as the failure of the measure of the path integral to be invariant under a given symmetry transformation. Let us consider the action $S[\Psi, \mathcal{G}]$, which is a functional of the fields $\Psi$ and background (gravitational) fields $\mathcal{G}$, such that it is invariant under the following linear transformation
\begin{equation}
\delta S = \frac{\delta S}{\delta \Psi} \delta \Psi + \frac{\delta S}{\delta \mathcal{G}} \delta \mathcal{G} = 0 \, ,
\label{app2.inft}
\end{equation}
where $\frac{\delta S}{\delta \mathcal{G}}$ is the densitized energy-momentum tensor. Here $\mathcal{G}$ represents the fields of the gravitational background, i.e. $\mathcal{G} = {h^{\mu \nu}, \tau^{\mu}, A_{\mu}}$ and $\tau_{\mu}$ for the NC background. On the shell of the equations of motion for $\Psi$, the first term on the right hand side of Eq.~(\ref{app2.inft}) vanishes, and the second term provides the classical conservation equation for the energy-momentum tensor
\begin{equation}
\frac{\delta S}{\delta \mathcal{G}} \delta \mathcal{G} = 0 \, .
\label{app2.set}
\end{equation}
Eq.~(\ref{app2.set}) represents Eqs.~(\ref{sf.diff}) and (\ref{sf.scale}). 
The quantum theory is described by the path integral
\begin{equation}
Z = \int \mathcal{D}\Psi e^{i S[\Psi,\mathcal{G}]} \, ,
\label{app2.pi}
\end{equation}
 whose measure involves only the quantum fields $\Psi$. The path integral is invariant under a given symmetry transformation of $\Psi$ provided
\begin{equation}
\int \mathcal{D}\Psi' e^{i S[\Psi',\mathcal{G}]}  = \int \mathcal{D}\Psi e^{i S[\Psi,\mathcal{G}]} \,.
\label{app2.pisym} 
\end{equation}
The effect of infinitessimal changes to the Jacobian and the action will provide the anomalous Ward identity. Considering Eq.~(\ref{app2.inft}), we have the following change in the action
\begin{align}
S[\Psi',\mathcal{G}] &= S[\Psi,\mathcal{G}] + \frac{\delta S}{\delta \Psi} \delta\Psi \notag\\
&= S[\Psi,\mathcal{G}] - \frac{\delta S}{\delta \mathcal{G}} \delta\mathcal{G} \, .
\label{app2.ds}
\end{align}
We also have the unitary transformation of the field $\Psi$, which can be written as
\begin{equation}
\Psi' = U \Psi = e^{i J} \Psi \, ,
\end{equation}
where $J$ is the Jacobian of the transformation. With this, the change in the functional Jacobian $\mathcal{I}$ (for a single bosonic field $\Psi$) is given by
\begin{equation}
\text{Det} U = e^{\text{Tr}\, \text{ln}\, U } \approx  1 + i\text{Tr} J 
\label{app2.jac}
\end{equation}
Using Eqs.(\ref{app2.ds}) and (\ref{app2.jac}) in Eq.~(\ref{app2.pisym}) now leads to the anomalous Ward identity
\begin{equation}
\left\langle \frac{\delta S}{\delta \mathcal{G}} \delta\mathcal{G}\right\rangle_{\Psi} = \left\langle\text{Tr} J \right\rangle_{\Psi} \quad \, ,
\label{app2.fdet}
\end{equation}
where $\left\langle \cdots \right\rangle_{\Psi}$ denotes the path integral average with respect to the variable $\Psi$. Thus the classical conservation equation is violated and results in an anomaly which is given by the functional trace of the Jacobian.
The trace is taken at the same point in spacetime, resulting in the presence of $\delta(0)$. Hence the trace of the Jacobian in Eq.~(\ref{app2.fdet}) is ill defined and requires regularization. As first demonstrated by Fujikawa \cite{Fujikawa:1979ay}, one can regulate using a positive definite operator $\mathcal{R}$ in the following way
\begin{equation}
\text{An} =  \lim_{M \to \infty}\text{Tr} J e^{-\frac{\mathcal{R}}{M^2}} =  \lim_{M \to \infty} \int d^n x \int d^n y J(x,y) e^{-\frac{\mathcal{R}(x)}{M^2}} \delta^n(x-y) \, ,
\label{app2.fuj}
\end{equation}
where the mode expansion for the functional trace in the last equality has been made for a scalar field. In Eq.~(\ref{app2.fuj}), $\text{An}$ denotes the candidate anomaly, not all of whose terms comprise the true anomaly. Only those terms for which a counterterm in the action cannot be provided will comprise the true anomaly. 

While this prescription is known to work, specific properties of the resultant gravitational anomalies depends on the choice of regulator. In the next subsection, we will consider how the regulators used in this paper agree with the Pauli-Villars scheme.

\subsection{Regulators} \label{app2.cr}
We will now present the arguments provided in \cite{Hatsuda:1989qy} which uses Pauli-Villars (PV) regularization to infer the corresponding Jacobian transformation and Regulator for Fujikawa's approach.
Let us consider the following action involving a collection of quantum fields $\Psi$
\begin{equation}
\mathcal{L}_{\Psi}= \frac{1}{2}\Psi^T \mathbf{T} \mathcal{Q} \Psi \, ,
\label{app2.act}
\end{equation} 
where for the purposes of this paper it will be suffice to assume that $\mathcal{Q}$ is any symmetric operator of mass dimension 2.  The superscript $T$ denotes transposition, while the symmetric matrix $\mathbf{T}$ in general depends on the background fields. Eq.~(\ref{app2.act}) is invariant under a certain symmetry transformation which we denote as
\begin{equation}
\delta_K \Psi = K \Psi \, .
\label{app2.sym}
\end{equation}
We now introduce the PV fields $\chi$, which are massive fields with the same statistics as $\Psi$, but with a different path integral definition to introduce a minus sign in one-loop graphs. Thus the Lagrangian is
\begin{align}
\mathcal{L}_{PV} &= \mathcal{L}_{\chi} + \mathcal{L}_{M} \notag\\
&= \frac{1}{2}\chi^T \mathbf{T}\mathcal{Q} \chi + \frac{1}{2} M^2 \chi^T \mathbf{T} \chi \, ,
\label{app2.pv}
\end{align}
where we have $M^2$ in the mass term due to $\mathcal{Q}$ in Eq.~(\ref{app2.act}) being a mass dimension 2 operator. The path integral is defined as
\begin{equation}
\int \mathcal{D}\chi e^{i \chi^T A \chi} = (\text{det} A)^{\frac{1}{2}}
\end{equation}
While we are considering only one copy of the PV fields, in general several copies are needed to cancel all possible one-loop divergences. The invariance of Eq.~(\ref{app2.act}) is now extended to the massless part of the PV action \footnote{Strictly speaking, we can have $\delta_{K'} \chi = K' \chi$, but in this case $K'$ must be such that the Jacobians of the fields $\chi$ and $\Psi$ cancel out. Here for simplicity, we have referred to $K'$ as $K$.}
\begin{equation}
\delta_K \chi = K \chi \, ,
\label{app2.chvar}
\end{equation}
such that the violation of symmetries, if any, can only arise due to the mass term. Under the transformation Eq.~(\ref{app2.chvar}) the mass term of the PV Lagrangian becomes
\begin{equation}
\delta_K \mathcal{L}_{M} = \delta_K \mathcal{L}_{PV} = \frac{1}{2}M^2\chi^T\left(\mathbf{T}K + K^T\mathbf{T} + \delta \mathbf{T} \right)\chi \, .
\label{app2.cdm}
\end{equation} 
Eq.~(\ref{app2.cdm}) can now be used to compute the anomaly due to the PV regulated path integral
\begin{align} 
\text{An}_K &= -\lim_{M\to \infty} \text{Tr}\left[\frac{1}{2}M^2 \left(\mathbf{T}K + K^T\mathbf{T} + \delta \mathbf{T} \right) \left(\mathbf{T}M^2+ \mathbf{T}\mathcal{Q}\right)^{-1}\right]\notag\\
&=-\lim_{M\to \infty} \text{Tr}\left[\left(K + \frac{1}{2} \mathbf{T}^{-1}\delta \mathbf{T}\right) \left(1 + \frac{\mathcal{Q}}{M^2}\right)^{-1} \right] \, ,
\label{app2.pvr}
\end{align}
where we could replace $K^T\mathbf{T}$ with $\mathbf{T}K$ since $\mathbf{T}$ and $\mathbf{T}\mathcal{Q}$ are symmetric.
From Eqs.(\ref{app2.fuj}) and Eq.~(\ref{app2.pvr}), we can identify the Jacobian and the regulator to be used in Fujikawa's approach as
\begin{equation}
J = K + \frac{1}{2}\mathbf{T}^{-1} \delta \mathbf{T} \, , \qquad \qquad \mathcal{R} = \mathcal{Q}
\label{app2.regjac}
\end{equation}
%
\subsection{Fujikawa regulators for non-relativistic field theories} \label{app2.app3}
While the comparison of PV regularization with that of the regulated trace in Fujikawa's approach has led to Eq.~(\ref{app2.regjac}), certain aspects of the calculation in the PV scheme are not present in Eq.~(\ref{app2.fuj}). Here we address the domain of integration of $\omega$ needed in the regulator to represent a non-relativistic one-loop calculation. Specifically, we will now argue that the correct regulated trace to be used in the Fujikawa approach to gravitational anomalies for non-relativistic theories should be
\begin{equation}
\lim_{M \to \infty} \text{Tr} J = \lim_{M \to \infty} \int \limits_{0}^{\infty}  \frac{d\omega}{2 \pi} \int \limits_{-\infty}^{\infty} \frac{d^2k}{(2 \pi)^2} e^{-i \omega t} e^{ikx}\left[J(x) e^{\frac{\mathcal{R}}{M^2}} \right] e^{i \omega t} e^{- ikx} \, .
\label{app2.fuj2}
\end{equation}
We recall that while one-loop effects in relativistic field theories involve pair creation and annhilation processes, vaccuum polarization effects, charge renormalization and mass renormalization, such processes are absent at one-loop for non-relativistic field theories~\cite{Bergman:1991hf,Bergman:1993kq}. The reason for this is that we can either have the forward time or the retarded time propagator. 
To understand what happens in the non-relativistic case let us first consider the Schr\"odinger field in $2+1$ dimensions. We perform the following mode expansion in terms of non-relativistic plane waves
\begin{align} 
 \Phi(x) &\sim e^{i\omega t - i k x} \notag\\
\Phi^*(x) & \sim e^{-i\omega t + i k x}
\label{an.me}
\end{align}
Given the action of $\mathcal{R}$ on $\Phi$ in the regulated trace and the mode expansion Eq.~(\ref{an.me}), we now want to determine what the range of the $\omega$ integral should be in order to represent the one-loop calculation. While we do not have access to the full Schr\"odinger propagator on curved backgrounds, it will suffice to consider the flat space operator to determine the nature of the $\omega$ integral. Taking $\mathcal{R} = i \partial_t + \frac{\nabla^2}{2}$, the propagator $G(x,t)$ satisfies
\begin{equation}
\left(i \partial_t + \frac{\nabla_x^2}{2}\right)G(x,x'; t, t') = \delta(t-t')\delta^2(x-x') \, ,
\label{app.prop}
\end{equation}
With the Fourier transform we have the following integral 
\begin{equation}
 G(x;t) = - \int \limits_{-\infty}^{\infty}\frac{d \omega}{2 \pi} \int \limits_{-\infty}^{\infty} \frac{d^2k}{(2 \pi)^2} \frac{e^{i \omega t - ikx}}{\omega + \frac{k^2}{2}}
\label{prop.four}
\end{equation}
This integral can be evaluated by choosing a pole \emph{either} in the upper half plane ($\omega \ge 0$) or the lower half plane ($\omega \le 0$). This freedom allows us to choose either the forward or retarded propagator. Given Eq.~(\ref{prop.four}) and the usual choice of the forward propagator for particles, this requires choosing the pole in the upper half plane
\begin{equation}
 G(x;t) = - \int \limits_{-\infty}^{\infty}\frac{d \omega}{2 \pi} \int \limits_{-\infty}^{\infty} \frac{d^2k}{(2 \pi)^2} \frac{e^{i \omega t - ikx}}{\omega + \frac{k^2}{2} - i\epsilon}
\label{prop.fo2}
\end{equation}
We can now readily integrate to find
\begin{equation}
G(x;t) = - \frac{\Theta(t)}{t}e^{-\frac{ix^2}{2t}}
\label{prop.result}
\end{equation} 
Since we will always consider the forward propagator for particles, we could have simply performed the integration over $\omega$ in Eq.~(\ref{prop.fo2}) from $0$ to $\infty$ without affecting the result. As the Fujikawa approach is meant to convey the one-loop calculation with this propagator for particles, we will perform our calculation in Fujikawa's approach with the regulator provided in Eq.~(\ref{app2.fuj2}).  
%
\section{BCH expansion terms} \label{app.C}
It will be convenient to introduce the following definitions
\begin{align}
\widetilde{\Gamma}^i &= \Gamma^i - 2im v^i \, , \notag\\
G^m &= 2 h^{-\frac{1}{4}}\partial^m h^{\frac{1}{4}} \, ,\notag\\
C^{ij} &= \Delta h^{ij} \, , \notag\\
D^{lij} &= \partial^lh^{ij} \, , \, D_l^{\phantom{l}ij} =\partial_lh^{ij} \, , \notag\\
E^{ij} &= \Delta C^{ij} + 2im D^{lij}\partial_l \mathcal{C} + \left(\Delta + G^m\partial_m\right)G^lD_l^{\phantom{l}ij} + G^l\partial_lC^{ij} - 2D^{lij}\partial_l\left(h^{-\frac{1}{4}}\Delta h^{\frac{1}{4}}\right) \, , \notag\\
H^{lij} &= \partial^l C^{ij}+  \Delta D^{lij} + D^{nij} \partial_n \widetilde{\Gamma}^l + \partial^l(G^mD_m^{\phantom{m}ij}) +G^mA_m^{\phantom{m}lij} -D_m^{\phantom{m}ij}\partial^mG^l \, , \notag\\
\Theta^{i j m n} &= D_k^{\phantom{k}ij} D^{kmn} \ , \notag\\
A^{i j m n} &= \partial^i D^{j m n} \, \, , \, \,A_{ij}^{\phantom{i}\phantom{i}mn} = \partial_iD_j^{\phantom{j}mn} \, ,\notag\\
B^{ijmn} &= - 2 \Theta^{i j m n} + 2 \left(A^{i j m n} + A^{j i m n}\right) \, ,
\label{app3.def}
\end{align}
where $v^i$ and $\mathcal{C}$ are as in Eq.~(\ref{an.milne}) and Eq.~(\ref{an.regc}) respectively. Then for $A = - k_i k_j h^{ij}$ and $B= \frac{i }{M} k_i \left(\widetilde{\Gamma}^i - 2 \partial^i  - G^i \right)+\frac{1}{M^2}\left(\Delta - i m \mathcal{C}  + h^{\frac{1}{4}}\Delta h^{-\frac{1}{4}} + G^l \partial_l\right)$ the BCH terms which describe $E$ in Eq.~(\ref{an.bch}) can be expressed as
\begin{align}
\left[A,B\right] &= -\frac{2 i}{M} k_i k_j k_m D^{mij} + \frac{1}{M^2} k_i k_j \left(C^{i j} + 2 D^{lij} \partial_l +  G^l D_l^{\phantom{l}i j} \right) \notag\\
\left[A,\left[A,B\right]\right] &= \frac{2}{M^2}k_i k_j k_m k_n \Theta^{ijmn} \notag\\
\left[B\left[A,B\right]\right] &= -\frac{4}{M^2}k_ik_jk_nk_mA^{mnij} -\frac{2i}{M^3}k_ik_jk_m\left(B^{ijml}\partial_l + H^{lij}\right)\notag\\
&\qquad \qquad \qquad \qquad + \frac{1}{M^4}k_ik_j\left(E^{ij} + B^{ijmn}\partial_m\partial_n +2 H^{lij}\partial_l\right) \notag\\
\left[A,\left[A,\left[A,B\right]\right]\right] &= 0 \, , \,~~~~\left[A,\left[A,\left[A,\left[A,B\right]\right]\right]\right] = 0 \, , \,
~~~~\left[B,\left[A,\left[A,\left[A,B\right]\right]\right]\right] = 0\notag\\
\left[A,\left[B,\left[A,B\right]\right]\right] &= -\frac{2i}{M^3}k_ik_jk_mk_nk_l B^{ijmp}\partial_ph^{nl} \notag\\
&\qquad \, + \frac{1}{M^4}k_ik_jk_mk_n\left(2 H^{lij}D_l^{\phantom{l}mn} + B^{ijlp}(A_{lp}^{\phantom{l}\phantom{p}mn} + 2D_l^{\phantom{l}mn}\partial_p)\right) \notag\\
\left[B,\left[B,\left[A,B\right]\right]\right] &= \frac{4}{M^4}k_ik_jk_mk_n\left[\left(B^{ijmp}D_p^{\phantom{p}nl}-2\partial^lA^{mnij} - \partial^nB^{ijml}\right)\partial_l -\partial^nH^{mij} \right. \notag\\&\left. \qquad \qquad \quad  \qquad \qquad - \Delta A^{mnij} - \frac{1}{2}B^{ijml}\partial_l\widetilde{\Gamma}^n \right] +\frac{8i}{M^3}k_ik_jk_mk_nk_l\partial^nA^{mlij}\notag\\
\left[A,\left[A,\left[B,\left[A,B\right]\right]\right]\right] &= \frac{2}{M^4} k_ik_jk_mk_nk_lk_kB^{ijpq}D_p^{\phantom{p}mn}D_q^{\phantom{q}lk} \notag\\
\left[A,\left[B,\left[B,\left[A,B\right]\right]\right]\right] &= \frac{4}{M^4} k_ik_jk_mk_nk_pk_q \left(B^{ijmr}D_r^{\phantom{r}nl}-2\partial^lA^{mnij} - \partial^nB^{ijml}\right)D_l^{\phantom{l}pq} \notag\\
\left[B,\left[B,\left[B,\left[A,B\right]\right]\right]\right] &= \frac{16}{M^4} k_ik_jk_mk_nk_pk_q \partial^p\partial^q A^{mnij}\notag\\
\left[B,\left[A,\left[B,\left[A,B\right]\right]\right]\right] &= -\frac{4}{M^4} k_ik_jk_mk_nk_pk_q \partial^p\left(B^{ijqr}D_r^{\phantom{r}mn}\right)
\label{app3.bch}
\end{align}
The free derivatives contained in the BCH terms above, and thereby in $E$, are needed in computing $E^2 \, , E^3$ and $E^4$ in Eq.~(\ref{an.exp}). With all expansions taken into consideration, we can drop the free derivative terms to arrive at Eq.~(\ref{an.exp2}). Only the terms $\mathcal{B}_2$ and $\mathcal{B}_4$ lead to non-trivial results following symmetric integration. By using $\partial_{\alpha}h^{ij} = -2\Gamma^{(i}_{\alpha k}h^{j) k} \, ; \alpha = (0,i)$, the terms contained in $\mathcal{B}_2$ are, order by order, given by
\begin{align}
k^0 &: -im\mathcal{C} \notag\\
k^2 &: -\frac{1}{2}k_ik_j \left(C^{ij} + \widetilde{\Gamma}^i \widetilde{\Gamma}^j  - 2 \partial^i \widetilde{\Gamma}^j\right)\notag\\
k^4 &: \frac{1}{3}k_ik_jk_mk_n \left(\Theta^{ijmn} + 2 A^{ijmn} \right) - k_ik_jk_mk_n\widetilde{\Gamma}^i D^{jmn} \notag\\
k^6 &: -\frac{1}{2}k_ik_jk_mk_nk_lk_k D^{lij}D^{kmn} \, .
\label{app3.b2}
\end{align}
Using these terms in Eq.~(\ref{an.symint}) results in the expression of Eq.~(\ref{an.b2}). The terms involved in $\mathcal{B}_4$ are considerably more involved and comprise the following
\begin{align}
k^0 &: -\frac{1}{2}\left(m^2\mathcal{C}^2 + im\Delta \mathcal{C}\right) \notag\\
k^2 &: k_ik_j\left[im\left(\frac{2}{3}\left(D^{lij}\partial_l\mathcal{C} + \partial^i\partial^j\mathcal{C} \right) +\frac{1}{2}\mathcal{C}\left(C^{ij} + \widetilde{\Gamma}^i\widetilde{\Gamma}^j \right) - \partial^i(\widetilde{\Gamma}^j\mathcal{C})\right) \right.\notag\\
&\qquad \qquad \qquad \qquad \qquad \qquad \qquad\left. -\frac{1}{6}\Delta C^{ij} -\frac{1}{2} \widetilde{\Gamma}^i\Delta \widetilde{\Gamma}^j + \frac{1}{3} \left(\Delta \partial^i \widetilde{\Gamma}^j + \partial^i \Delta \widetilde{\Gamma}^j \right) \right] \notag\\
k^4 &: k_ik_jk_mk_n \left[im\left( \mathcal{C}\left(\widetilde{\Gamma}^iD^{lmn}-\frac{1}{3}\left(\Theta^{ijmn} + 2 A^{ijmn}\right)\right) - D^{jmn}\partial^i\mathcal{C}\right) + \frac{1}{6}\left(\partial^i \partial^j C^{mn} +  \partial^i \Delta D^{jmn}\right) \right. \notag\\
& \qquad \qquad \qquad - \frac{1}{3}\widetilde{\Gamma}^i\left(\partial^j C^{mn} + \Delta D^{jmn}\right) -\frac{1}{2} \partial^i(D^{lmn} \partial_l \widetilde{\Gamma}^j)+ \frac{2}{3}\widetilde{\Gamma}^i D^{lmn} \partial_l\widetilde{\Gamma}^j + \frac{1}{6}\Delta\left(\Theta^{ijmn} +  A^{ijmn}\right)\notag\\
&  \qquad \qquad \qquad \quad + C^{ij}\left( \frac{1}{8}C^{mn} +  \frac{1}{4}\widetilde{\Gamma}^m\widetilde{\Gamma}^n -  \frac{1}{2}\partial^m\widetilde{\Gamma}^n\right)- \frac{1}{12}D_{l}^{\phantom{l}ij}\left(\Delta D^{jmn} + D^{lmn}\partial_l\widetilde{\Gamma}^j -2 \partial^l C^{mn}\right)  \notag\\
&  \qquad \qquad \qquad \qquad +\frac{1}{12} B^{ijml}\partial_l\widetilde{\Gamma}^n + \frac{1}{24}\left(\widetilde{\Gamma}^i\widetilde{\Gamma}^j\widetilde{\Gamma}^m\widetilde{\Gamma}^n - B^{ijlp}A_{lp}^{\phantom{l}\phantom{p}mn} \right) -\frac{1}{2}D^{mij}\Delta \widetilde{\Gamma}^n -\frac{1}{3}D^{lij}\partial_l\left(\partial^m \widetilde{\Gamma}^n\right) \notag\\
& \left. \qquad \qquad \qquad \qquad  \qquad +\frac{1}{2} \left(\partial^i \widetilde{\Gamma}^j\right)\left(\partial^m \widetilde{\Gamma}^n\right) + \frac{2}{3}\widetilde{\Gamma}^m\partial^i\partial^j \widetilde{\Gamma}^n - \frac{1}{3}\partial^i\partial^j \partial^m \widetilde{\Gamma}^n -\frac{1}{2}\widetilde{\Gamma}^i\widetilde{\Gamma}^j\partial^m\widetilde{\Gamma}^n \right]\notag\\
k^6 &: k_ik_jk_mk_nk_lk_k \left[\frac{1}{2}im\mathcal{C}D^{lij}D^{kmn} -\frac{1}{3}\left(\partial^lC^{ij} + \Delta D^{lij}\right)D^{kmn} -\frac{1}{6}C^{ij}\left(\Theta^{mnlk} + 2 A^{lkmn} - 3 \widetilde{\Gamma}^lD^{kmn}\right)\right. \notag\\
&\qquad  \qquad \qquad \quad -\frac{1}{9}\partial^l\partial^k \left(\Theta^{ijmn} + \frac{11}{5}A^{ijmn}\right)  +(\Theta^{ijmn} + 2 A^{ijmn})\left(\frac{1}{3}\partial^l\widetilde{\Gamma}^k -\frac{1}{6}\widetilde{\Gamma}^l\widetilde{\Gamma}^k\right) - \widetilde{\Gamma}^lD^{mij}\partial^n\widetilde{\Gamma}^k\notag\\
& \qquad \qquad \qquad \quad \quad + D^{rij}\partial_r\left(\frac{1}{2}\widetilde{\Gamma}^l\partial_rD^{kmn}+\frac{2}{3}D^{kmn}\partial_r\widetilde{\Gamma}^l -\frac{1}{6}\partial_r(\Theta^{lkmn}+ A^{lkmn})\right) + \frac{1}{6}\widetilde{\Gamma}^l\widetilde{\Gamma}^k\widetilde{\Gamma}^m D^{nij}\notag\\
& \qquad \qquad \qquad \qquad \quad +\frac{1}{3}\left(\widetilde{\Gamma}^l\partial^k(\Theta^{ijmn} + A^{ijmn}) - \partial^l(D^{rij}\partial_rD^{kmn}) + 2 D^{mij}\partial^l\partial^k\widetilde{\Gamma}^n\right)\notag\\
&\qquad \qquad \qquad \qquad \qquad +\frac{1}{60}D_{r}^{\phantom{r}mn}\left(B^{ijrs}D_{s}^{\phantom{s}lk} - B^{ijlr}\widetilde{\Gamma}^k - 2 \partial^r A^{lkij}\right) \notag\\
&\left. \qquad \qquad \qquad \qquad \qquad \qquad +\frac{1}{15}\left(B^{ijmr}\partial_rD^{nlk} +\frac{7}{6}\partial^l(B^{ijkr} D_{r}^{\phantom{r}mn})\right)\right]\notag\\
k^8 &: k_ik_jk_mk_nk_lk_kk_pk_q\left[\frac{1}{18}\left(\Theta^{ijmn}\Theta^{lkpq} + 4 A^{ijmn}A^{lkpq}+ 4 \Theta^{ijmn}A^{lkpq}\right) - \frac{1}{12}B^{ijmr}D_r^{\phantom{r}nl}D^{kpq}  \right.\notag\\
& \qquad \qquad\qquad \qquad \quad +\frac{1}{2}\left(C^{ij}D^{pmn}D^{qlk} + D^{pij}D^{rmn}\partial_rD^{qlk}\right) +\frac{1}{3}\partial^p\left(\Theta^{ijmn} + A^{ijmn}\right) D^{qlk} \notag\\
& \left. \qquad \qquad\qquad \qquad \qquad +\left(\frac{1}{4} \widetilde{\Gamma}^i \widetilde{\Gamma}^j -\frac{1}{2}\partial^i \widetilde{\Gamma}^j\right) D^{pmn}D^{qlk} -\frac{1}{3}\left(\Theta^{ijmn} + 2 A^{ijmn}\right)\widetilde{\Gamma}^pD^{qlk}\right]\notag\\
k^{10} &: k_ik_jk_mk_nk_lk_kk_pk_qk_rk_s\left[\frac{1}{6}\left(\widetilde{\Gamma}^r D^{sij}D^{pmn}D^{qlk} - \left(\Theta^{ijmn} + 2 A^{ijmn} \right)D^{rpq}D^{slk}\right)\right] \notag\\
k^{12} &: \frac{1}{24}k_ik_jk_mk_nk_lk_kk_pk_qk_rk_sk_uk_v D^{rij}D^{smn}D^{ulk}D^{vpq} \notag \\
\label{app3.b4}
\end{align}
%

\end{document}